\documentclass[aip,amsmath,amssymb,reprint]{revtex4-1}

\usepackage{graphicx}
\usepackage{dcolumn}
\usepackage{bm}

\newcommand{\Ez}{$E_z$}
\newcommand{\current}{\textit{j}}
\newcommand{\aapp}{$a_{\text{app}}$}
\newcommand{\ahop}{$a_{\text{hop}}$}
\newcommand{\astruc}{$a_{\text{struc}}$}
\newcommand{\aea}{$a_{E_A}$}
\newcommand{\Dpara}{${D}_{\parallel}$}

\newcommand{\Dortho}{${D}_{\perp}$}

\newcommand{\moby}{$\mu$}
\newcommand{\muLi}{$\mu_{\text{Li}^+}$}

\newcommand{\Li}{$\text{Li}^+$}
\newcommand{\TFSI}{$\text{TFSI}^-$}
\newcommand{\meter}{\text{m}}

\newcommand{\mus}{$\mu\text{s}$}
\newcommand{\opeo}{$\text{O}_{\text{PEO}}$}

\newcommand{\ie}{\textit{i.e.}}
\newcommand{\eg}{\textit{e.g.}}

\newcommand{\etal}{\textit{et al.}}

\newcommand{\angstrom}{\mbox{\normalfont\AA}} 

\usepackage{graphicx} 
\usepackage{subcaption} 

\usepackage{float}  
\usepackage{placeins}  

\usepackage{siunitx} 
\usepackage[]{physics} 

\newcommand{\Lim}[1]{\raisebox{0.5ex}{\scalebox{0.8}{$\displaystyle \lim_{#1}\;$}}} 

\begin{document}

\preprint{APS/123-QED}

\title{Exploring nonlinear ion dynamics in polymer electrolytes from the perspective of hopping models}

\author{Alina Wettstein}
\author{Diddo Diddens\textsuperscript{\textdagger} }
\author{Andreas Heuer}%
 
\affiliation{Institut für Physikalische Chemie, Universit{\"a}t M{\"u}nster, D-48149 M{\"u}nster, Germany \\  andheuer@uni-muenster.de \\
\textsuperscript{\textdagger}Helmholtz Institute M{\"u}nster (IMD-4), Forschungszentrum J{\"u}lich GmbH, M{\"u}nster, 48149, Germany}%

\date{\today}


\begin{abstract}

Relevant information about the nature of the dynamics of ions in electrolytes can be obtained by studying the nonlinear dependence on an applied electric field. Here we use molecular dynamics (MD) simulations to study the field effects for a polymer electrolyte, i.e. a mixture of PEO with Li-TFSI salt, for a range of different temperatures and salt contents. Specifically, the effects of the electric field on the current and the diffusivities parallel and orthogonal to the electric fields are analyzed.  It is argued that the nonlinear effects in the weak-field regime provide information about the nature of the disorder.  In contrast, the nonlinear effects in the high-field regime allow one to extract effective hopping distances. They are in close agreement with typical nearest-neighbor length scales obtained from detailed structural analysis and are hardly dependent on the salt content.  Furthermore, from the study of the temperature dependence in the high-field regime, effective barrier heights can be determined, which in this regime decrease linearly with increasing field.  The disappearance of the effective barriers, estimated by linear extrapolation, occurs close to the fields where the MD simulations start to be numerically unstable. Finally, the interpretation is supported by a comparison with analytically known solutions of  disordered hopping models. \\
\\
Please note that the contents of this manuscript are a revised version of chapter 5 from Alina Wettstein\textquotesingle s dissertation \cite{wettstein2022dissertation}; this revised version is now refined with perspectives and contributions from co-authors.

\end{abstract}


\maketitle


\section{\label{sec:introduction} Introduction}

Faced with a progressively diverse range of lithium-ion battery applications, optimization strategies have been tailored to the specific operational situations and resulted in a great diversity of battery architectures. \textit{E.g.}, once a niche interest, miniaturized batteries have received growing attention as they are vital to the implementation of standalone micro-devices such as medical implants, micro-sensors or next-generation smart cards  \cite{ferrari2015latest,ni2020three,jetybayeva2021recent}.  
To date, a safe micro-sized battery design relies on thin-film solid electrolytes which can be classified into organic and inorganic types, each with its advantages and disadvantages. Extensive research effort has been invested in inorganic glass electrolytes because of the relatively facile processability of thin films and particularly high thermal and electrochemical stabilities \cite{jetybayeva2021recent,roling2008field}. Polymer electrolytes, on the other hand, have emerged as a promising organic alternative in terms of achievable ionic conductivity which is up to two orders of magnitude higher than for the yet more stable glass options \cite{jetybayeva2021recent}.
Applying a working voltage of up to 5\,V on thin ($\lesssim 1\,\mu\meter$) electrolyte films results in large cell-internal electric fields \cite{roling2008field,ponce2021analysis}. Computational research on diverse  electrode-electrolyte setups has provided quantitative estimates on the arising internal potential gradients, \eg{} by evaluating the Poisson potential resulting from ionic concentration profiles \cite{Matse2020,Jorn2013}, and it has been found that in narrow regions close to the interface layered ion structures, that extend over several \angstrom{} or nm into the electrolyte, may evoke strong, very localized fields of up to $\mathcal{O}$(1\,V/nm) \cite{Matse2020,Jorn2013,Stuve2011,lepley2015modeling}.
To understand and improve the \Li{} transport kinetics in these systems, it is thus crucial to investigate the properties of field-induced dynamic nonlinearities.  

Progress has been achieved in the description and interpretation of nonlinear transport phenomena in the field of solid ion conductors, connecting insights  from experiment, simulation, and theory \cite{roling2008field,kunow2006nonlinear,heuer2005nonlinear,heuer2014physical}. When the nonlinear effects were interpreted as resulting from the hopping of particles in a simple sinusoidal system, the observed hopping distances (up to 4.3\,nm) turned out to be much larger  than hopping distances, estimated from structural properties. Therefore, they have been denoted apparent hopping distances \aapp{}. This indicates that the difference between \aapp{} and the actual hopping distances requires the presence of disorder so that one may hope to learn about the underlying disorder of the underlying energy landscape. 

The \Li{} transport in a polymer melt is more complicated since it is a superposition of two kinds of hopping events which are associated with different effective ranges \cite{brooks2018atomistic,maitra2007cation,diddens2010understanding}: additionally to lithium's co-diffusion with bound polymer segments, the \Li{} ions explore the polymer backbone via jumping along consecutive coordination sites, \ie{}, intra-chain hopping (1). Since the polymer chains are of finite length long-range \Li{} transport strongly depends on so-called inter-chain transfers (2), meaning \Li{} affiliates with another polymer strand and acquires a new range of motion. However, it should be stressed that despite these different aspects \Li{} dynamics in a polymeric host can be modeled through elementary jumps which therefore establishes yet a close connection to the concepts used for glassy electrolytes. Therefore, it may be interesting to compare the nonlinear conductivity spectra of polymeric systems with those of solid ion conductors. Indeed, for polymeric systems nonlinear conductivity spectra, \ie{}, the coefficients $\sigma_1$ and $\sigma_3$, have been determined for a PEO/$\text{LiClO}_4$ ($r\,=\,$ \Li{}/EO\,unit\,=\,0.1) electrolyte  above its glass transition point \cite{tajitsu1996linear}. Interestingly, in that work the resulting apparent hopping distances are reported to be even larger (of the order of 100\,nm) than those obtained for glassy solid ion conductors, indicative of the presence of even stronger nonlinear effects.

In contrast to the solid ion conductor the polymer chains undergo continuous rearrangement which means that the lithium ions experience a dynamically evolving energy landscape. This implies that when \Li{} is blocked by a high local barrier on its migration path, the constant segmental relaxation of the polymer makes it very likely that this high barrier will be replaced by a smaller one after some time. Qualitatively, the flexible polymer lattice thus suppresses dead-end pathways and effectuates that \Li{} motion is restrained by effectively lower barriers. Thus, one might expect that the effective disorder of the energy landscape is smaller.

Here we study the electric field-induced nonlinear ion dynamics in polymer melts, \ie{}, poly(ethylene oxide) (PEO) electrolytes at different lithium-bis(trifluourmethane)sulfonimide (LiTFSI) salt concentrations,  using classical MD simulations. 
Recent electrophoretic NMR experiments reported field-dependent ion mobilities in PEO/LiTFSI electrolytes and mentioned the possibility of a field-induced ordering of the polymer host matrix that would provide structural pathways for efficient ion transport \cite{Rosenwinkel2019}. 
Even though our previous investigation of these mixtures revealed strongly nonlinear dynamics above \Ez{}$\gtrsim 0.1$\,V/nm as well as a coiled-to-stretched transformation of the polymer chains upon field application, we could elucidate that the enhanced transport properties are not of a structural origin but can be attributed solely to field-tilted activation barriers \cite{wettstein2021polymer}. 
We show to what extent one can understand and analytically describe the nonlinear field dependence of various dynamic observables. By evaluation the numerical results we present an improved understanding of the underlying properties of the energy landscape, relevant for the lithium dynamics. In particular, we show that different pieces of information can be gained from individually studying the electric low-field and high-field regime, respectively.

\section{\label{sec:hopping} Nonlinear transport: theoretical background}

\subsection{General}

A key observable for probing nonlinear ion transport is the drift current density \current{}. Its information content is equivalent to that of the electrophoretic mobility \moby{} which characterizes how the ion drift velocity $\nu \propto j$ scales with the external electric field \Ez{}, i.e. $\mu\,=\,\nu\,/\,E_z$.
Assuming \current{} and \moby{} to be analytical functions of \Ez{}, their field dependencies can be expressed each by a polynomial series
\begin{equation} 
    \begin{alignedat}{3}
    &j(E_z) &&= \sigma_1 \cdot E_z &&+ \sigma_3 \cdot E_z^{\,3} +  \sigma_5 \cdot E_z^{\,5} + \dots\\
    \qquad &\mu(E_z) &&= \mu_0 &&+ \mu_2\cdot E_z^{\,2} + \mu_4\cdot E_z^{\,4} + \dots 
\end{alignedat}
\label{eq:power_series_j_mu}
\end{equation}
Since a reversal of the field direction must result in a reversal of \current{} without affecting its absolute value, the current's field dependence involves exclusively odd power terms of \Ez{}. Analogous symmetry reasons imply an even power series for \moby{}. The coefficients, which capture the linear and nonlinear field contributions, can be measured both in experiment and simulation \cite{kunow2006nonlinear,roling2008field,mattner2014frequency,heuer2014physical,heuer2005nonlinear}. 

\begin{figure*}
\includegraphics[width=0.9\textwidth]{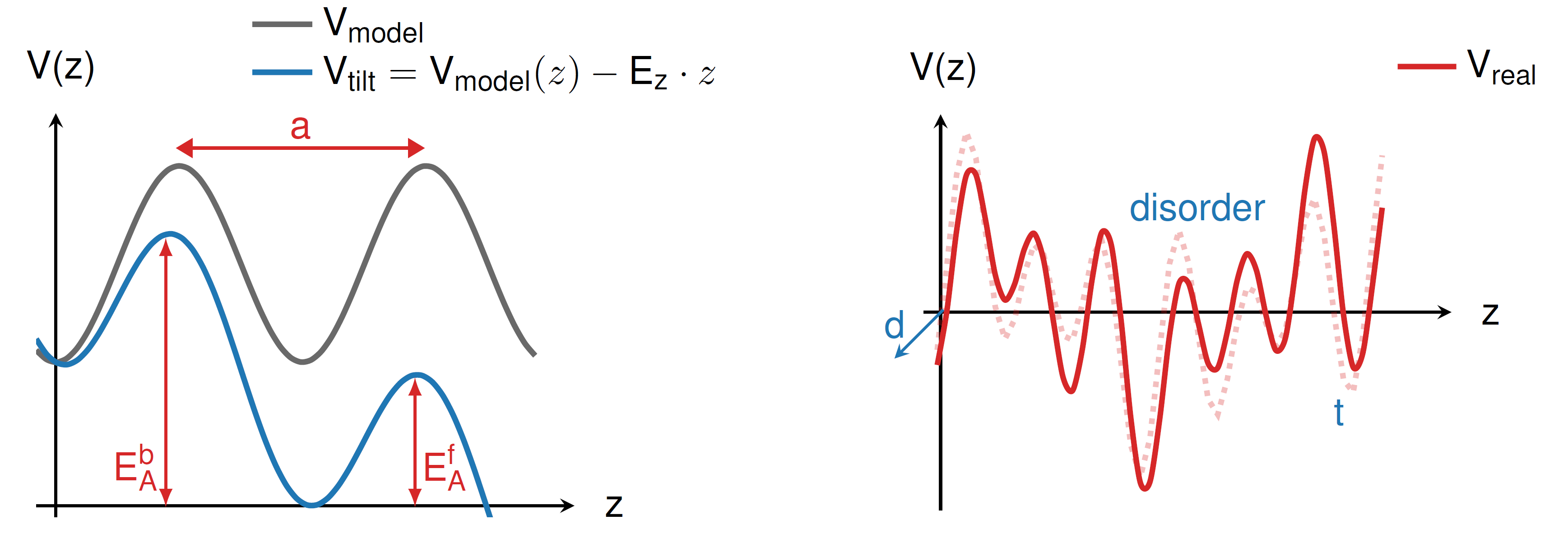}
\caption{\label{fig:1d_model_potentials}Left: Sketch of 1D model potential $V_{\text{model}}(z)$ (gray) which serves as a starting point for the theoretical analysis of nonlinear hopping dynamics. The sinusoidal profile experiences a tilt (blue) upon field application which raises the barrier heights $E_A^b$ for backward jumps and lowers $E_A^f$ for forward jumps. The discrete lattice sites are a distance $a$ apart. Right: Schematic indications (blue) in which respects $V_{\text{real}}$ (red) deviates from the simplified assumptions of $V_{\text{model}}$ (left): dynamics occur in dimensions $d>1$ on a disordered potential energy surface. In soft matter systems, the potential barriers may furthermore change over time $t$ (dotted red).}
\end{figure*}

\subsection{Sinusoidal energy landscapes}

In an attempt to rationalize the physical meaning of the nonlinearities, single-particle hopping models were devised where the mobile charge carrier is pictured to perform thermally activated jumps between neighboring sites in sinusoidal energy landscape. Figure \ref{fig:1d_model_potentials} (left) illustrates for a sinusoidal profile, where vacant particle sites are a distance $a$ apart, how the application of an electric field tilts the potential energy surface: while the activation energy ${E}_{{A}}^{{f}}$ for forward motion is lowered, a relatively increased ${E}_{{A}}^{{b}}$ raises an energetic penalty for reverse dynamics. To be strictly accurate, the field does not modify exclusively the barrier heights but also the positions of the local minima and maxima. As subtly apparent in the sketched example, an increasing field strength shortens the gap between a potential minimum and subsequent maximum to less than $a/2$ for forward jumps while lengthening the distance to more than $a/2$ for backward jumps. In the regime of sufficiently small \Ez{}, where the effects of position shifting can be neglected, one can predict the nonlinear enhancement of ion motion in field direction analytically from the difference in the forward and backward jumping rates $\Gamma_{{f}}$ and $\Gamma_{{b}}$ which are proportional to $  \exp[( -\beta \cdot (E_A^0\ \mp  q \cdot (a/2) \cdot E_z)]$. One obtains
\begin{equation} 
    j  \propto  \big[ \Gamma_f -\Gamma_b\big]  \\
   \equiv \dfrac{\Gamma_0}{2}\,\cdot\,\sinh\left( \beta\,\cdot\,q\,\cdot\, \frac{a}{2}\,\cdot\,E_z \right) 
\label{eq:sinh_relation_j_1D_basis}
\end{equation}
which in the low-field regime can be written via a Taylor expansion as
\begin{equation} 
        j\stackrel{\text {Taylor}}{=}  \underbrace{\frac{\Gamma_0}{2}\beta q \frac{a}{2}}_{\sigma_1} \cdot E_z \,+\, \underbrace{\frac{\Gamma_0}{2}\dfrac{1}{6}\left(\beta q\frac{a}{2}\right)^3}_{\sigma_3}\cdot E_z^3 + ...  .
\label{eq:sinh_relation_j_1D}
\end{equation}
Here $\beta$ is the inverse temperature, $q$ denotes the particle's charge and $\Gamma_0$ a rate factor. The prefactors of the first terms in dependence on \Ez{} reflect the polynomial coefficients in Equation \ref{eq:power_series_j_mu}. Even more, it is now possible to extract the hopping distance $a$ from knowledge of the linear and the cubic term:
\begin{equation}
a = \sqrt{\dfrac{24\,(k_BT)^2\,\sigma_3}{q^2\sigma_1}} .
\label{eq:a_app}
\end{equation}

Although the hopping model serves as a physical basis to motivate, \eg{}, the current being an odd function of field strength which must hold due to the fundamental principle of symmetry, experimental measurements of $a$ strongly indicate that the nonlinear dynamics in solid ion conductors are insufficiently described by this ansatz: the typically measured values of $a\,\approx\,15-40\,$\angstrom{} dramatically exceed structural distances $a_{\text{struc}}$ between neighboring ion sites which are usually only in the order of a few \angstrom{} \cite{heuer2005nonlinear,barton1996electric,isard1996high,banhatti2001structure,kunow2006nonlinear}. If coarse-graining the dynamics in form of an effective lattice was a viable approach, one would expect $a_{\text{struc}}$ to give a rough indication of $a$. Therefore, we denote the lengthscale, obtained via Equation \ref{eq:a_app} as $a_{\text{app}}$. 

\FloatBarrier

\subsection{Randomly oriented hopping directions}

\begin{figure}[htb!]
    \centering
    \begin{subfigure}[t]{0.45\linewidth}
        \includegraphics[width=\textwidth]{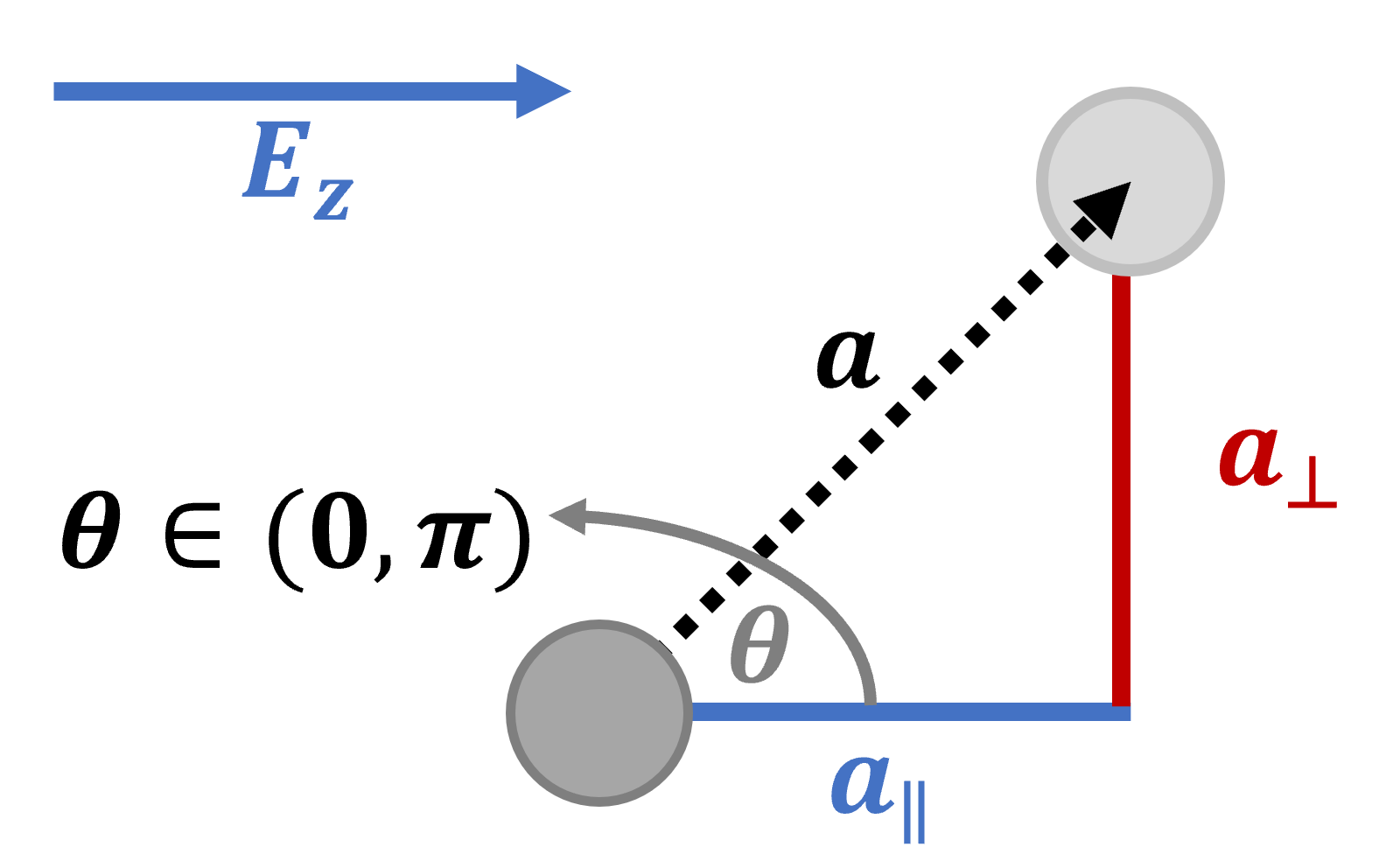}
    \end{subfigure}
  \caption{A particle can perform hops at any angle $\theta$ relative to the direction of the electric field $E_z$. The effective hopping distance $a$ (black) can accordingly be decomposed into a parallel $a_{\parallel} = \cos(\theta)\cdot a$ (blue) and an orthogonal $a_{\perp}= \sin(\theta)\cdot a$ (red) part. } 
  \label{fig:scheme_a_para_ortho_decomposition}
\end{figure}

The previous analysis, but also many other hoping models have in common that they consider dynamics on a regular lattice which is constructed such that ions may hop either parallel or orthogonal to the field direction. To go beyond this simplification, it has been pointed out that in 3D disordered   materials ion jumps will have different orientations toward the field vector, which alters the mathematical description \cite{kunow2006nonlinear}.  The energy barriers for, \eg{}, forward motion $E_A^f$ are only lowered with respect to the ion's displacement $a_{\parallel} \leq a$ in field direction, \ie{} $E_A^f = E_A^0 - q\cdot a_{\parallel}\cdot E_z / 2$. It is therefore necessary to decompose a particle hop into parallel and orthogonal components as visualized in Figure \ref{fig:scheme_a_para_ortho_decomposition}, and to perform an angular average over all possible jump angles $\theta$ when deriving the field dependency of any dynamic observable.
   As a consequence one observes, quite counterintuitively, field-enhanced dynamics perpendicular to the field as reflected by field-dependent orthogonal diffusivity \Dortho{}.

Specifically, one needs to correct the forward $\Gamma_f$ and backward $\Gamma_b$ hopping rates with the correct energy gain or penalty that the ion receives from a jump $a(\theta)$ in the tilted energy landscape. 
As can be gathered from Figure \ref{fig:scheme_a_para_ortho_decomposition}, the contribution from an overall jump distance $a$ is given by $a_{\parallel}\,=\,a\cdot\cos(\theta)$. 
In order to consider all possible jump orientations $0\leq\theta\leq\pi$, where $\theta=0$ denotes a field-aligned forward jump with $a_{\parallel}=a$, and $\theta=\pi$ corresponds to a field aligned reverse jump with $a_{\parallel} = -a$, Equation \ref{eq:sinh_relation_j_1D_basis} is transformed into an integral:
\begin{equation}
    \begin{alignedat}{1}
    &\mu(E_z) = \dfrac{\langle a_{\parallel}(\Delta t) \rangle}{E_z\cdot \Delta t} \\
             &= \quad \dfrac{1}{E_z\cdot \Delta t}\,\cdot\,\dfrac{1}{2}\int_0^{\pi} a\cos(\theta)\Gamma_0\Delta t\cdot e^{u\cdot\cos(\theta)}\,\,\sin(\theta)d\theta,
    \end{alignedat}
    \label{eq:ansatz_derivation_mu_u}
\end{equation}
with  $u = q\cdot\beta\cdot a/2 \cdot E_z$. Please note that this ansatz considers the scenario of only a single particle hop $a(\cos{\theta})$ occurring within the time interval $\Delta t$. Naturally, this assumption is only reasonable in the limit of short times $\Delta t \rightarrow 0$. Since the ion dynamics follow a random walk, though, and can thus be treated as a Markovian process, the field-dependence expected for $\Delta t\rightarrow \infty$ can be retrieved from this short-time behavior. The integral is straightforwardly solved using the substitution $x=u\cdot\cos(\theta)$, which yields:
\begin{equation}
    \mu(E_z) = \dfrac{a}{E_z}\,\cdot\,\Gamma_0\left[\dfrac{\cosh{u}}{u} - \dfrac{\sinh{u}}{u^2} \right].
    \label{eq:exact_solution_mu_u}
\end{equation}
where the field dependence is contained in $u$. 
In analogy to Equation \ref{eq:ansatz_derivation_mu_u}, the nonlinear behavior of the parallel diffusivity \Dpara{}(\Ez{}), \ie{}, the diffusive dynamics of ion species $i$ in the moving coordinate frame of the respective drift motion $D_{\parallel\,\,i} = \lim \limits_{t\rightarrow\infty} \dfrac{\langle (z_j(t)-z_j(0))^2 \rangle_{j \in i} -  \nu_i^2 \cdot t^2}{2 \cdot t}$ 
and \Dortho{}(\Ez{}) can now be derived from the angular average of the second moments $\langle r_{\parallel}^2\rangle$ and $\langle r_{\perp}^2 \rangle$ according to (steps of partial integration not shown):
\begin{equation}
    \begin{alignedat}{2}
    &D_{\parallel}(E_z) = \dfrac{\langle a_{\parallel}^2(\Delta t) \rangle}{ 2 \Delta t} \\
    &= \quad\dfrac{1}{2\Delta t}\,\cdot\, \dfrac{1}{2}\int_0^{\pi} a^2\cos^2(\theta)\Gamma_0\Delta t\cdot e^{u\cdot\cos(\theta)}\,\,\sin(\theta)d\theta \\
   &=\quad \dfrac{a^2\,\Gamma_0}{2} \,\cdot\,\left[ \dfrac{(u^2+2)\sinh{u}}{u^3} - \dfrac{2\cosh{u}}{u^2} \right]
    \end{alignedat}
\label{eq:exact_solution_d_para_u}
\end{equation}
as well as
\begin{equation}
    \begin{alignedat}{2}
    &D_{\perp}(E_z) = \dfrac{\langle a_{\perp}^2(\Delta t) \rangle}{ 4 \Delta t}  \\
    &= \quad\dfrac{1}{4\Delta t}\,\cdot\, \dfrac{1}{2}\int_0^{\pi} a^2\sin^2(\theta)\Gamma_0\Delta t\cdot e^{u\cdot\cos(\theta)}\,\,\sin(\theta)d\theta \\
   &=\quad \dfrac{a^2\,\Gamma_0}{2} \,\cdot\,\left[ \dfrac{\cosh{u}}{u^2} - \dfrac{\sinh{u}}{u^3} \right].
    \end{alignedat}
\label{eq:exact_solution_d_ortho_u}
\end{equation}

One may wonder why the square of the first moment $\langle r_{\parallel}(\Delta t)\rangle^2$ is not relevant for the derivation of \Dpara{}, given that the MD data processing requires subtraction of the drift motion from the measured parallel mean squared displacement $\text{MSD}_{\parallel}$. Here we explore that the dynamics in this simple hopping model is strictly Markovian so that the diffusion constant can be evaluated in the short-time limit. However, in this limit contribution of the ballistic term to the diffusivity is proportional to $t^2/t = t$ and thus strictly disappears. 

Now we are in the position to deduce \aapp{} values separately for \moby{}, \Dpara{} and \Dortho{} from second order Taylor expansions $\propto \lambda_0 + \lambda_2\cdot E_z^2$ in the low-field regime (see Equations \ref{eq:A5_taylor_mu}-\ref{eq:A5_taylor_d_ortho} in Appendix \ref{appendix:Taylor_expansion}).  Straightforward calculations yield
    \begin{equation}
     \begin{alignedat}{2}
        a_{\text{app}}^{\,\mu} = &a_{\text{app}}^{\,D_{\perp}} = \sqrt{\dfrac{40(k_BT)^2 \lambda_2}{q^2 \lambda_0}} \\ 
        &a_{\text{app}}^{\,D_{\parallel}} = \sqrt{\dfrac{40(k_BT)^2 \lambda_2}{3 \cdot q^2 \lambda_0}} .
     \end{alignedat}
     \label{eq:lambda_mapping}
    \end{equation}

\subsection{Analysis of disordered models in the low-field regime}

In order to better understand the experimentally observed large values of \aapp{}, disordered energy landscapes have been analyzed, see the sketch   in Figure \ref{fig:1d_model_potentials} (right). It turns out that the nonlinear properties are significantly influenced by the characteristics of the underlying energy landscape \cite{roling2002hopping,rothel2010theoretical,heuer2014physical} which, in general, may be even time-dependent.

By means of analytical calculations of multidimensional models with disorder, supported by numerical modeling,  it was possible to identify two limiting scenarios giving rise to either enhanced or retarded net dynamics \cite{heuer2014physical}:
\begin{itemize}
    \item[(i)] If a highly disordered system features only a small number of low-energy sites, the ions are originally captured by these minima. When an external field is employed, however, the ions may escape from such trapping regions and are pushed to states of higher energy where they are more mobile and add to an increased ion flux, represented by an increase of $\sigma_3$ and thus \aapp{}. However, as explicitly shown in Appendix \ref{appendix:PEL_disorder_app_d_mu}, the value of \aapp{} may slightly differ, if obtained from analysis of the field-dependent mobility or diffusivity.
    \item[(ii)] In the opposing case, where high-energy sites are the minority, the transport in the linear response regime is governed by the low-energy sites. However, upon increasing field strength, some ions will be shifted to sites which are blocked by an adjacent high-energy site. This may be inaccessible to the ion. The effectively barrier-trapped ions cannot contribute to the macroscopic current, leading to a decreasing nonlinearity which may even end up with $\sigma_3\,<\,0$. Numerically, it turns out that this behavior is even observed for an equal number of low- and high-energy sites
\end{itemize}

These examples demonstrate also from the modelling side that the true hopping distance $a$ in general cannot be extracted from the ratio between the nonlinear and linear response coefficients. This is even more extreme for $\sigma_3\,<\,0$ which would imply an imaginary value of $a_{\text{app}}$ (see Equation \ref{eq:a_app}), which obviously is not physically meaningful. This is why it was termed an \textit{apparent} hopping distance \aapp{}. 
The above examples also show that the nature of nonlinear effects are closely related to the nature of the energy landscape, governing the ion transport.

\subsection{Analysis of disordered models in the high-field regime}

\begin{figure*}
    \centering
    \begin{subfigure}[t]{0.45\linewidth}
        \includegraphics[width=\textwidth]{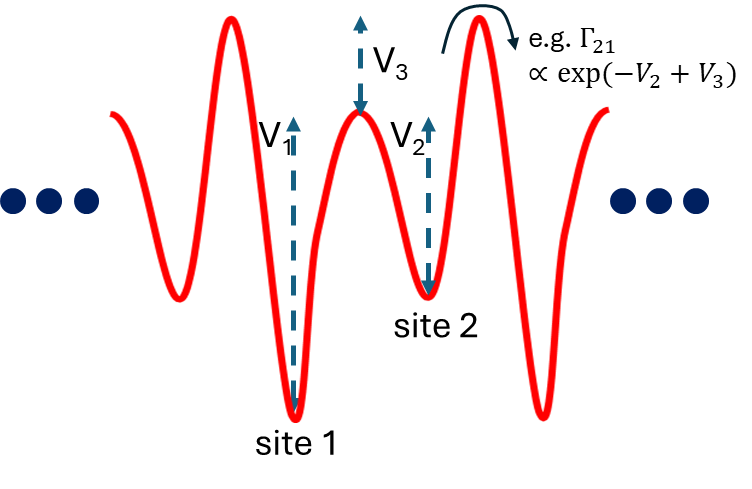}
    \end{subfigure}
  \caption{Schematic representation of disordered energy landscape (here sketched in 1D) with two different sites as analyzed in Ref. \cite{mattner2014frequency}.}
  \label{fig:A5_scheme_adapted_from_Mattner2014}
\end{figure*}

Mattner \etal{} \cite{mattner2014frequency} investigated nonlinear ion transport upon application of time-modulated electric fields within a single-particle hopping model in a disordered energy landscape of arbitrary dimension $d$ with two energy sites. To simplify the model, low and high energy sites are strictly alternating. A 1D sketch is shown in
Figure \ref{fig:A5_scheme_adapted_from_Mattner2014}. The results of the  analytically solvable model can be translated straightforwardly into the less complex scenario of a constant field \Ez{}. The general result, presented in Appendix \ref{appendix:hopping_model_analytical}, can be written as 
\begin{widetext}
\begin{equation}
    j = \exp( \beta q a E_z /2) \left [\frac{2\Gamma_1 \Gamma_2 \Gamma_3}{\Gamma_1 + \Gamma_2 \Gamma_3}+\exp( -\beta q a E_z/2  )\cdot f(\Gamma_1,\Gamma_2,\Gamma_3,d)+ {\cal O}(\exp(-\beta q a E_z  ))\right ] .
    \label{eq:A5_current_high_field_regime}
\end{equation}
\end{widetext}
with some algebraic function $f$ containing the energy parameters and the dimensionality of the model. Here the $\Gamma_i \equiv \exp(-\beta V_i)$ are dimensionless.

It turns out that for $d > 1$ one has a {\it decreasing} nonlinearity $\sigma_3$ with an increase in field strength in the low-field regime. This is consistent with the behavior discussed above since in this model the number of low-energy and high-energy sites is identical. 

Here, we use this model to discuss the limit of high electric fields, since we argue that some generally relevant behavior can be extracted. We concentrate on fields in positive directions. Of course, for reasons of symmetry, an analogous behavior is observed when inverting the direction of the field. In the high-field limit one obtains $ j(E_z) \propto  \exp( \beta q a E_z /2)$. This result is remarkable, since even for a disordered energy landscape the elementary hopping distance $a$, denoted \ahop{} in this work, can be extracted from the field dependence in the high-field regime. 

The emergence of the factor $\exp(\beta q a  E_z  /2)$ has a straightforward physical explanation. In the high-field regime the particle is only allowed to perform jumps in the direction of the field. Thus, for each transition the barrier is reduced by $q a  E_z  /2$, giving rise to 
this exponential factor in  Eq.\ref{eq:A5_current_high_field_regime}. Furthermore, due to the effective 1D dynamics, the dependence on dimensionality disappears.

The current in high-field regime can be rewritten as
\begin{equation}
    j = \frac{2}{\tau_1 + \tau_2}
    \label{eq:A5_current_high_field_regime2}
\end{equation}
where $\tau_i = \exp(\beta (B_i - q a E_z/2))$ is the average waiting time to stay in site $i$ and $B_i$ is the respective barrier to escape to the right side without the additional field-contribution. In the present example, one has $B_1 = \exp(-\beta V_1)$ and $B_2 = \exp(-\beta (V_2+V_3))$. Equation     \ref{eq:A5_current_high_field_regime2} can be rewritten as  $1/j = \langle \tau_i\rangle$. This expression is valid for the most general disordered energy landscape as can be seen, e.g., from the high-field limit of the solution of a 1D model with arbitrary disorder \cite{heuer2005nonlinear}. This expression results from the observation that in the high-field limit the total time it takes for a particle to move some distance is just the sum of the individual waiting times of the sequence of minima.

Furthermore, one may determine the apparent activation energy $-d \log(j) / d\beta$ in the high-field regime for different values of $E_z$.  In the high-field limit one obtains $const - q a E_z/2$ so that from analyzing the activation energy in dependence of $E_z$,  the hopping distance can be read again from the slope. To distinguish it from the previous analysis, we denote it as \aea{}. Within this model, it should be identical to \ahop{}.

 When applying these model results to the actual analysis of the simulated data in the high-field regime we have again to take into account the isotropic average of hopping directions. Here, the two approaches, yielding  \ahop{} and \aea{}, respectively, differ.
 In the first case, we have mapped the disordered system on a sinusoidal 1D energy landscape. This is justified because the same high-field limit $j \propto \exp( \beta q a E_z /2)$ is expected. Then, we have used Eqs.\ref{eq:exact_solution_mu_u}, \ref{eq:exact_solution_d_para_u}, \ref{eq:exact_solution_d_ortho_u} to fit the numerical data. In the route via the activation energy this mapping is not required, but we can directly calculate the apparent activation energy from the numerical data. This suggests that the determination of \aea{} is more robust.

\section{Simulation details}

Since the work draws to the simulation series presented in Ref. \cite{wettstein2021polymer}, the reader is referred to \textit{ibid.} for a detailed description of the system generation as well as equilibration protocol. 
The studied mixtures consist of 20 relaxed PEO chains which are each 27 monomers long. Varying amounts of Li\,TFSI salt, \ie{}, 36, 54, and 81 ion pairs, are embedded in the polymer matrix so that the salt-to-monomer ratio $r\,=\,$[Li]/[EO] spans over $r$\,=\,0.07, 0.10 to 0.15. 

In summary, the MD simulations were carried out using the GROMACS 2018.6 software package \cite{VanDerSpoel2005,Pall2015,Abraham2015,Berendsen1995}. The PEO interactions are modeled according to the standard OPLS-AA force field, while \Li{} and \TFSI{} are described by the OPLS-AA compatible CL\&P parametrization \cite{gouveia2017ionic,CanongiaLopes2012,JoseN.CanongiaLopes2004,Lopes2004,Shimizu2010}. 
The production runs were performed in the \textit{NPT} ensemble using a Parrinello-Rahman barostat \cite{Parrinello1981} to maintain atmospheric pressure and a Nos\'{e}-Hoover thermostat \cite{Nose1983,Nose1984,Hoover1985} to couple the system to a temperature of 423\,K if not mentioned otherwise. The trajectories are at least 300\,ns but up to 2\,\mus{} long to ensure that the diffusive regime is sampled with sufficient statistics.

Furthermore, this work extends the range of electric field strengths \Ez{} by a high-field regime $1.0<E_z\leq 2.0\,$V/nm. Interestingly, even larger field strengths of \Ez{}\,=\,3.0\,V/nm lead to unstable and crashing simulations, for which the subsequent analysis will provide a descriptive explanation. Lastly, simulations at different temperatures $400\,\text{K}\leq T \leq 440\,\text{K}$ were additionally run for the $r\,=\,$0.07 composition at various fields between 0.0 to 2.0\,V/nm. 

\section{Results and discussion}

\subsection{Structural properties}

The previous investigation revealed that the structural landscape in which \Li{} and \TFSI{} move is dramatically modified by the external field \cite{wettstein2021polymer}.It was found that the lithium ions, tightly bound to the polymer and migrating in field direction, pull the PEO chains into an elongated shape. Associated with these stretched chains, the crown ether-like wrapping of \Li{} is increasingly suppressed, resulting in lower \Li{}-monomer coordination numbers. An in-depth discussion of these results and their implications for the \Li{} dynamics is provided in Ref. \cite{wettstein2021polymer}.

\begin{figure*}
    \centering
    \begin{subfigure}[t]{0.45\linewidth}
        \includegraphics[width=\textwidth]{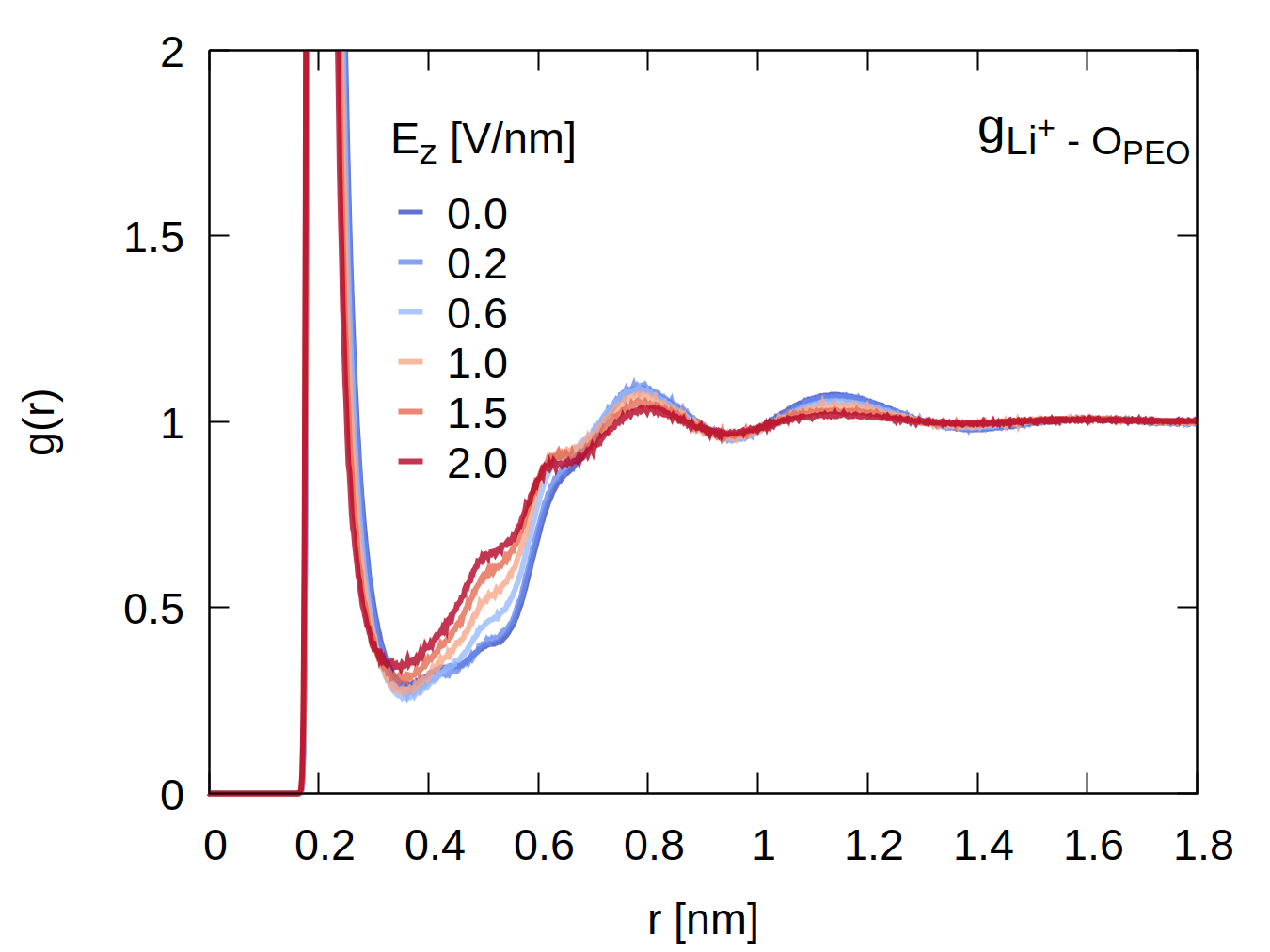}
        \caption{\label{fig:rdf_li_o_peo}}
    \end{subfigure}
    ~
    \begin{subfigure}[t]{0.45\linewidth}
        \includegraphics[width=\textwidth]{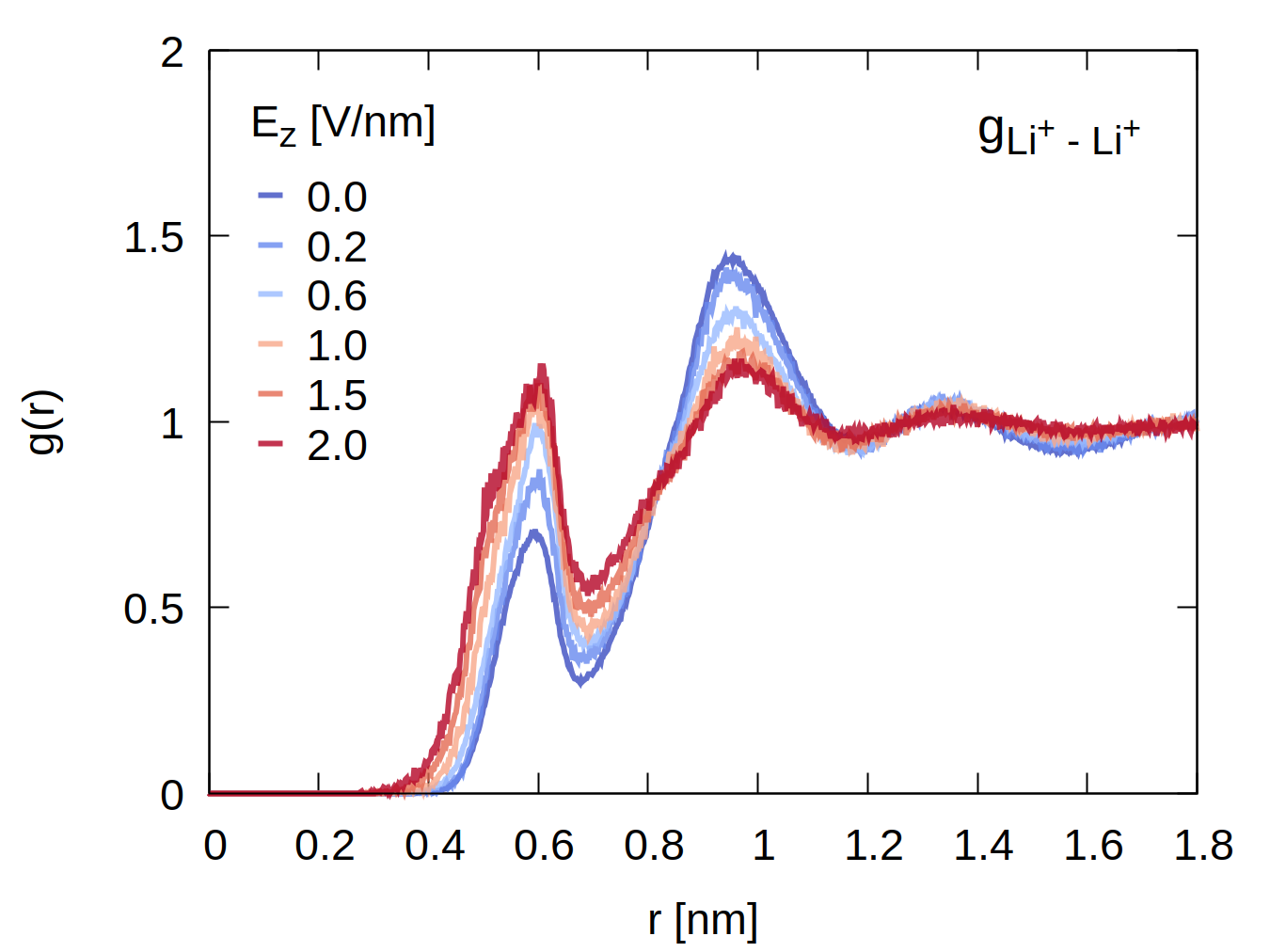}
        \caption{\label{fig:rdf_li_li}}
    \end{subfigure}
    ~
      \caption{Field dependence of radial distribution functions $g(r)$ for \Li{}-$\text{O}_{\text{PEO}}$ (a) and \Li{}-\Li{} (b) for the $r\,=\,0.07$ mixture. The data for the fields up to 1 V/nm are reprinted (adapted) with permission from  A. Wettstein, D. Diddens,  and A. Heuer, Polymer electrolytes in strong external electric fields: Modification of structure and dynamics, Macromolecules $\bold{54}$, 2256 (2021) \cite{wettstein2021polymer}. Copyright 2021 American Chemical Society.} 
  \label{fig:structural_hopping_distances}
\end{figure*}

Given that this study aims to rationalize the field-dependent ion dynamics through hopping models, the ion hopping distance \ahop{} constitutes a key observable. The radial distribution functions (RDF) $g(r)$ between, \eg{}, \Li{} and the polymer ether oxygens \opeo{} as well as other \Li{} emerge as especially interesting in this regard because they provide direct insight into the most probable next-neighbor distances. Figure \ref{fig:structural_hopping_distances} gives an overview of the arrangement of \Li{}-\opeo{} (a) and \Li{}-\Li{} (b) in the mixture for a broad spectrum of electric field strengths.
The field dependence of  \Li{}-\opeo{} RDF is shown in Figure \ref{fig:rdf_li_o_peo}. All curves exhibit a strong and sharp peak at a distance of 2\,\angstrom{} which marks lithium's first coordination shell. An increasing field strength does not affect the peak position, \ie{}, how close \Li{} and \opeo{} favorably approach each other, but steadily decreases the peak height. This field-induced weakening of the peak intensity is equivalent to the previously mentioned lower \Li{}-monomer coordination numbers, which are given by the integral of the RDF up to the first minimum position. 
Since the stretching of the polymer chains in field direction impedes an all-around complexation of \Li{} by consecutive \opeo{} of a single chain, adjacent but non-coordinating monomers are pushed into the region between first and second coordination shells. This is reflected in the increasingly pronounced shoulders developing at distances of 5 and 6\,\angstrom{}. 
Lastly, we note that the \Li{}-\opeo{} structuring is very localized and fades quickly after merely two repeat distances. From the \Li{} perspective, the polymer matrix does not exhibit ordered properties beyond distances of 1.2\,nm.

Figure \ref{fig:rdf_li_li} shows how the population of \Li{} around \Li{} responds to increasing field strengths. All curves display two distinctive peaks located at distances of about 6 and 9\,\angstrom{}. Interestingly though, the relative peak heights invert with increasing field strength. When no field is applied, the \Li{}-\Li{} RDF assumes values below one for distances below 7\,\angstrom{} and the second peak is the most pronounced feature which, on the one hand, demonstrates a general avoidance between distinct \Li{}. 
The well-defined peak located at 6\,\angstrom{}, on the other hand, indicates that rare structural configurations may assist a closer approach. A possible interpretation of the length scales could be that the spatial extent of the first solvation shell sets a threshold for how close the lithium ions would approach each other. As can be seen from the first minimum position of $g_{\text{Li}^+-\text{O}_{\text{PEO}}}(r)$, the primary \Li{} wrapping by \opeo{} is contained within $\approx$ 3\,\angstrom{}, which is compatible with the \Li{}-\Li{} peak signature at 6\,\angstrom{}.  
As an increasing field strength is applied, the intensity of the next-neighbor peak increases and its width broadens which means that the \Li{}-\Li{} distribution is progressively shifted towards shorter distances. 
Given the electrostatic repulsion between \Li{} pairs, this trend behavior is not intuitively understandable and likely the result of an intricate combination of effects. On a qualitative level, the field-induced structuring of the polymer host may serve as an explanation. The fact that \Li{} is continuously forced out of a nestled polymer environment and in return bound to an increasingly stretched PEO strand, makes the closer yet sufficiently distant next-neighbor site more accessible. It seems plausible that the repulsive interaction between adjacent \Li{} would then be screened by interstitial \TFSI{} slipping into the gaps. The observed concurrent increase of \Li{}-\TFSI{} coordination numbers is supportive of this interpretation. 

In summary, analysis of the typical spacing between \Li{} and \opeo{} as well as other \Li{}, respectively, demonstrates that the local environment around \Li{} is strongly ordered and provides us with a structural window \astruc{} for the dynamic \Li{} hopping distances \ahop{}. For one thing, one would anticipate \ahop{} to be larger than the characteristic size of the first coordination shell which is 2\,\angstrom{}$\leq$\ahop{}. The approximate distance to the second-next adjacent \opeo{} is indicated by the first shoulder in the \Li{}-\opeo{} RDF developing at $5\,$\angstrom{}. Then again, it would be unreasonable for \Li{} to jump further than the distance of 6\,\angstrom{} between neighboring \Li{}. 
\FloatBarrier

\subsection{Field-dependent transport properties}

Here we show the field-dependent values of  \moby{}, \Dpara{} and \Dortho{}  for \Li{}. According to our theoretical analysis they can be interpreted in the low- as well as in the high-field regime. To analyse the low-field regime we have fitted each curve by the second order Taylor expansion $\propto \lambda_0 + \lambda_2\cdot E_z^2$ and then used Eq.\ref{eq:lambda_mapping} to extract the respective values of \aapp{}. We have checked that in the regime  $E_z\leq 0.5\,$V/nm higher order terms can be neglected so that we restricted the fitting to this regime. In contrast, in the high-field regime we have taken the expressions in Eqs.\ref{eq:exact_solution_mu_u}, \ref{eq:exact_solution_d_para_u}, and \ref{eq:exact_solution_d_ortho_u}, respectively. This analysis also involves two adjustable parameters, namely the prefactors and the hopping distances $a_{\text{hop}}$, respectively. Since the values of $a_{\text{hop}}$ can be read off from the slope of the transport quantity (in logarithmic representation) in high-field limit the precise fitting range is not important (we have chosen $E_z\,\geq\,0.8\,$V/nm).

\begin{figure*}
    \centering
    \begin{subfigure}[t]{0.45\linewidth}
        \includegraphics[width=\textwidth]{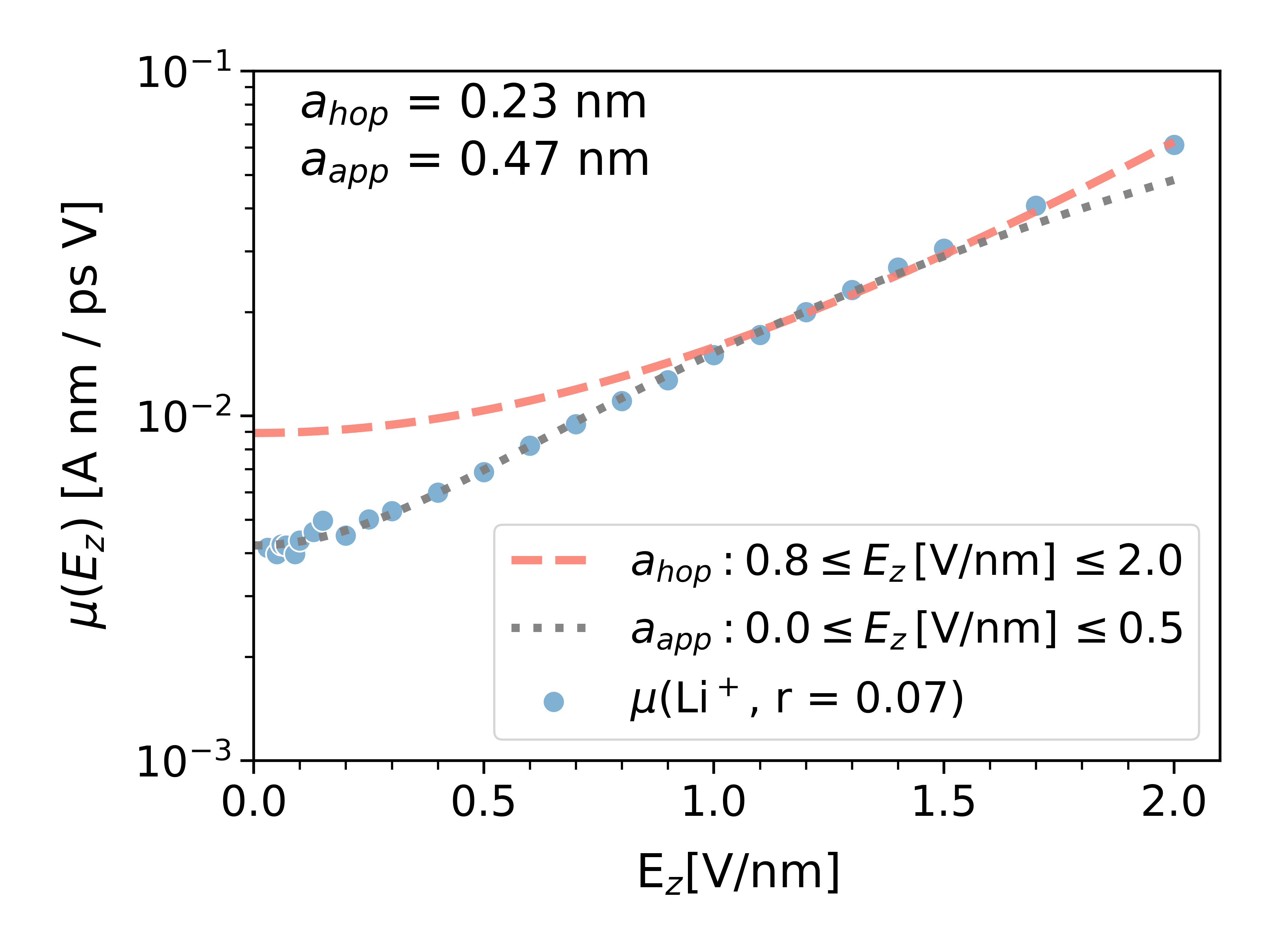}

        \label{fig:mobility_r_0_06_li}
    \end{subfigure}
    ~

    \begin{subfigure}[t]{0.45\linewidth}
        \includegraphics[width=\textwidth]{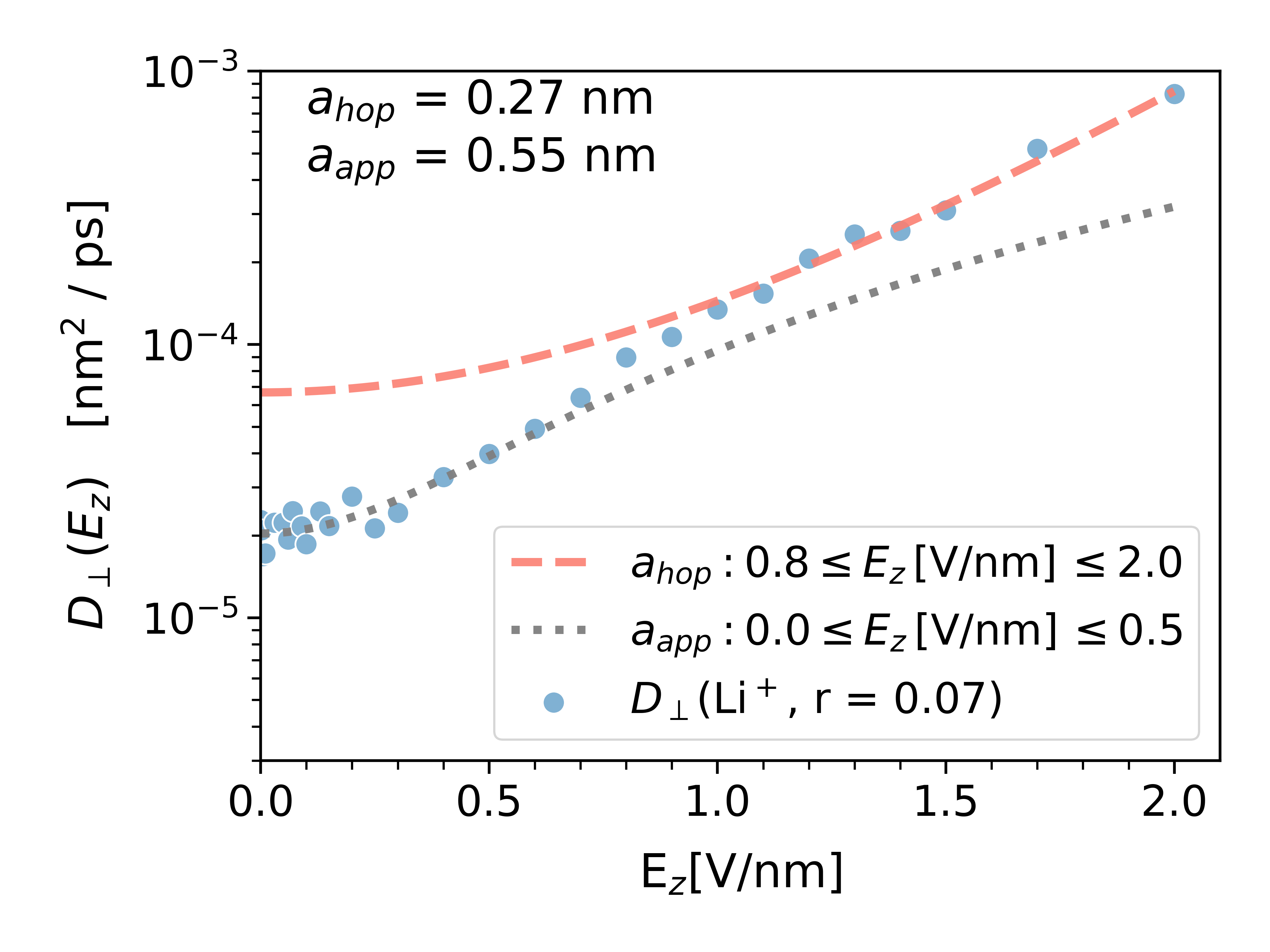}

        \label{fig:D_ortho_r_0_06_li}
    \end{subfigure}
    ~

    \begin{subfigure}[t]{0.45\linewidth}
        \includegraphics[width=\textwidth]{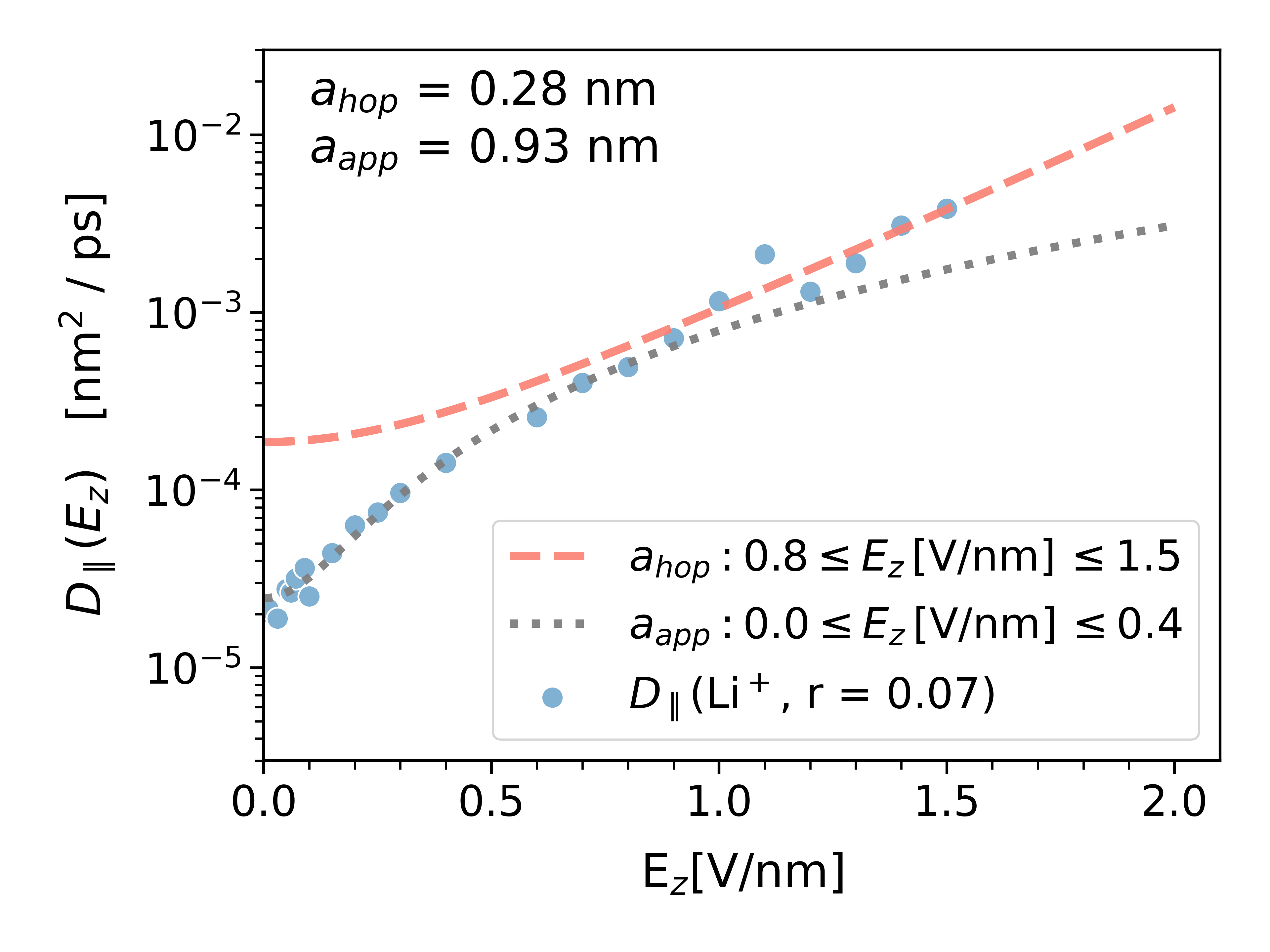}
 
        \label{fig:D_para_r_0_06_li}
    \end{subfigure}
    ~ 
     \caption{Field dependence of \moby{} (top), \Dortho{} (middle) and \Dpara{} (bottom) for \Li{}  for the $r\,=\,0.07$ mixtures. The data is fitted separately in the low-field (gray) and the high-field (red) regime. While the red fits employ the analytically derived relations, the gray fits are based on the corresponding Taylor expansion of the analytical description for the low-field regime. The thus obtained hopping distances \aapp{} (low fields) and \ahop{} (high fields) are listed. The data for the fields up to 1 V/nm are reprinted (adapted) with permission from  A. Wettstein, D. Diddens,  and A. Heuer, Polymer electrolytes in strong external electric fields: Modification of structure and dynamics, Macromolecules $\bold{54}$, 2256 (2021) \cite{wettstein2021polymer}. Copyright 2021 American Chemical Society.  } 
  \label{fig:analytic_fitting_data_low_high_field_regime}
\end{figure*}

At this point, the results are discussed exemplarily for the $r=0.07$ mixture while a summarized presentation including all concentrations will be given below. Figure \ref{fig:analytic_fitting_data_low_high_field_regime} shows an overview over the low-field regime fits (gray) to obtain \aapp{} and the high-field regime fits (red) to extract \ahop{}.
\FloatBarrier

We find that the apparent \Li{} hopping distances \aapp{}$\,\geq\, 4.7\,$\angstrom{} for all dynamic characteristics are larger than the typical next-neighbor distances of 2\,\angstrom{} that were estimated from the \Li{}-\opeo{} radial distribution functions. 
Interestingly, the largest value of nearly 1\,nm is associated with \Dpara{} which is consistent with the \aapp{} ranking observed for silicate glasses \cite{kunow2006nonlinear}.
However, apart from this similarity, the discrepancy between \aapp{} and the presumed \astruc{} turns out markedly lower in these polymer melts compared to literature reports on apparent \Li{} hopping distances in the range of several nanometers for glassy systems \cite{heuer2005nonlinear,barton1996electric,isard1996high,banhatti2001structure,kunow2006nonlinear}.  
As discussed in the introductory section, \aapp{} should not be interpreted as an average jump distance even though it carries the unit of a length. It rather encodes information about very general characteristics of the potential energy landscape explored by the ions. 
In analogy to the mechanistic understanding developed in Ref. \etal{} \cite{heuer2014physical}, the observation of \aapp{}$>$\astruc{} in this work could indicate that the energy landscape is best described to contain scattered trapping regions on an otherwise leveled surface (trapping regime). One could tentatively imagine the energetic traps on a local level as temporarily very stable polymer coordination motifs or, on a higher level of abstraction, as \Li{} being trapped with a distinct chain before transferring onto another one. If ion dynamics were modulated by very high energy barriers, on the other hand, one would observe $a_{\text{app}}^2<a_{\text{struc}}^2$ (barrier regime). Please note that the comparison is consciously described for the squared distances because, in principle, also negative nonlinear effects may arise in barrier-dominated energy landscapes. Our findings thus suggest an effective absence of high barrier blocking, which is consistent with the earlier presumed special features of polymer energy landscapes: The continuous relaxation of the polymer is equivalent to a temporal variation of the energy barriers which should ultimately pave the landscape traversed by the ions since a high confining barrier will soon reduce its height again, thus enabling ion passage.

The \Li{} hopping distances \ahop{}\,$\approx 2.5\,$\angstrom{} obtained from the high-regime fits of the three dynamic observables show reasonable agreement among themselves and also match the structurally determined distance to adjacent \opeo{}. A more careful view shows that for all salt concentrations \ahop{} is  approximately 20\% higher for $\mu$ as compared to \Dortho{}. Here we just mention that in our system two additional effects have to be taken into account. First, for high electric fields the polymer starts to stretch in the direction of the electric field which may yield a slightly anisotropic distribution of hopping directions. Second, this effects increases with increasing field strength, both increasing the actual mobility as compared to the theoretical expantion as expressed in Equation \ref{eq:exact_solution_mu_u}. The same effects would decrease \Dortho{} as compared to the isotropic case. Thus, it is not surprising that fitting procedure, based on the isotropic assumption ,yields slightly different values of \ahop{}. However, a closer analysis of these small differences is beyond the scope of this work.  

We would like to add that the high-field determination of \Dpara{} suffers from the fact that the MSD has to be evaluated in the moving coordination system. Thus, it has to be determined as the difference of two large numbers of similar size, reflecting the second moment and the squared first moment.  This numerical challenge gives rise to the relatively large scattering of the data in the high-field regime. 

\subsection{Activation energies}
\begin{figure}[H]
    \centering
    \begin{subfigure}[t]{0.9\linewidth}
        \includegraphics[width=\textwidth]{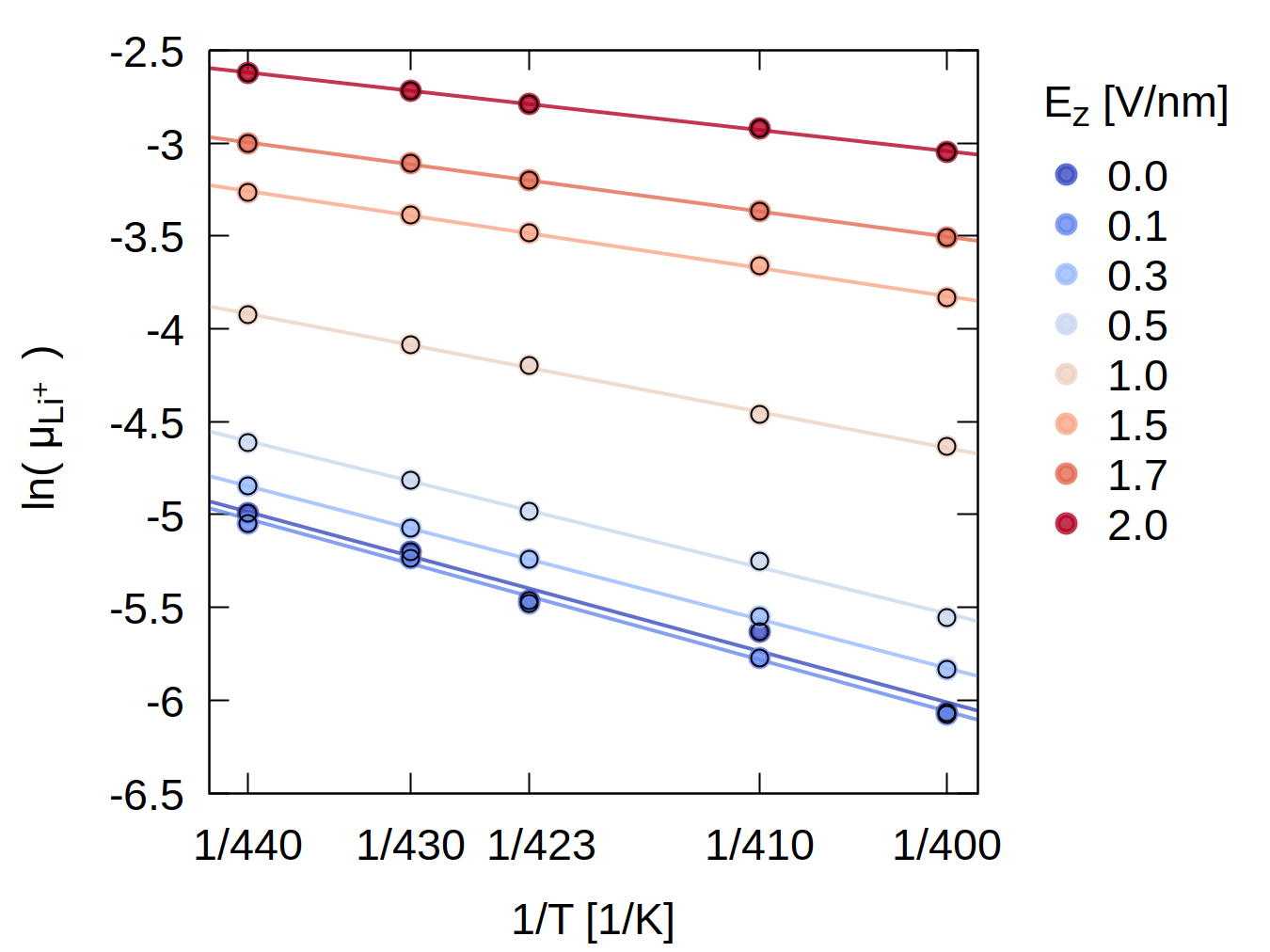}
    \end{subfigure}
  \caption{Arrhenius plot of \Li{} mobilities in the r\,=\,0.07 mixture over a broad range of electric field strengths \Ez{}. The straight lines are linear fits where the slope measures the field-dependent activation energy $E_A$.} 
  \label{fig:arrhenius_plot_lithium_mobility}
\end{figure}

While the preceding analysis mainly reflected the dynamic implications of the field-tilted energy landscape, the barrier tilting itself can be evidenced directly through temperature-dependent measurements, \eg{}, \moby{}(\Ez{}, $T$).
Figure \ref{fig:arrhenius_plot_lithium_mobility} shows an Arrhenius plot exemplary for \muLi{} at various field strengths. Within the investigated temperature range of 400\,K$\,\leq \,T \leq$\,440\,K the temperature dependencies of $\ln{\mu (E_z, 1/T)}$ and $\ln{D_{\perp} (E_z, 1/T)}$ are fitted by an Arrhenius-like relationship, thus extracting an activation energies $E_A$ from the linear slopes. Given the quite narrow temperature window, the value of  $E_A$ might be interpreted as an apparent activation energy in this temperature regime. We refrained from analyzing \Dpara{} due to the large numerical uncertainties (see above).

\begin{figure}[!htb]
    \centering
    \begin{subfigure}[t]{0.9\linewidth}
        \caption{\Li{} activation energies.}
        \includegraphics[width=\textwidth]{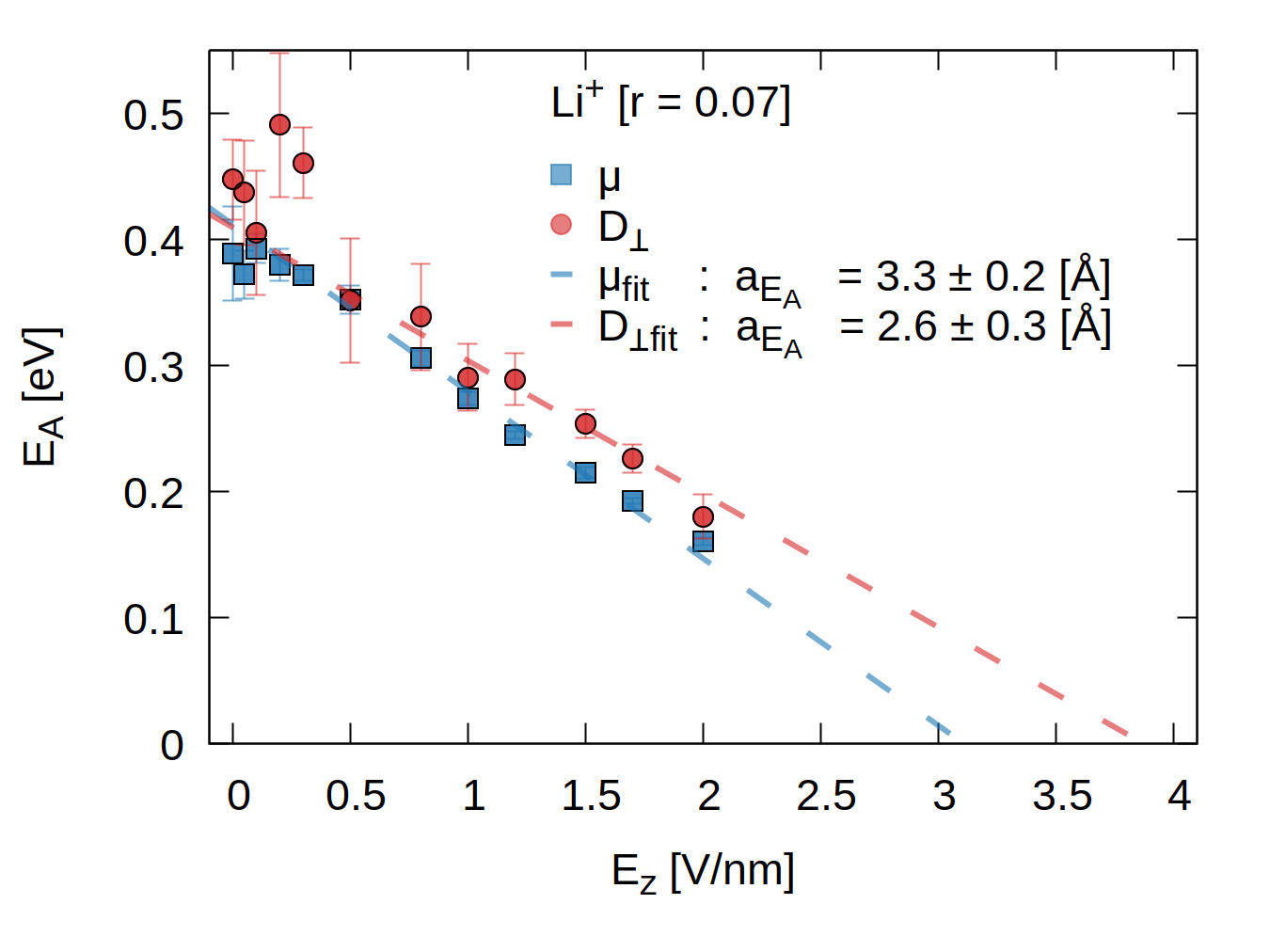}
        \label{fig:E_A_function_of_E_z_Li}
    \end{subfigure}

  \caption{\Li{} activation energies $E_A$ as a function of electric field \Ez{}. $E_A$ is derived from the temperature dependence of the mobilities \moby{} (blue) and the orthogonal diffusivities \Dortho{} (red). The linear relationship in the regime of high fields $E_z\,\geq\,0.5\,$V/nm yields a hopping distance $a_{\text{E}_\text{A}}$ according to $E_A = E_A^{\,0} - q\cdot a_{\text{E}_\text{A}}/2 \cdot E_z$.} 
  \label{fig:E_A_function_of_Ez}
\end{figure}

Figure \ref{fig:E_A_function_of_Ez} depicts the activation energies of drift motion and orthogonal diffusion as a function of \Ez{}. It shows that the energy barriers are lowered with increasing field strength as anticipated and, furthermore, that drift and diffusion dynamics exhibit very similar $E_A(E_z)$. 
We note that the $E_A$ values are in good agreement with recent MD simulation results \cite{brooks2018atomistic} as well as PFG NMR measurements by Gorecki \etal{} yielding $E_{A}^{\text{Li}^+}=\,$\SI{0.38}{\eV}  in a $r$\,=\,0.05 PEO electrolyte of high molecular weight ($N_{\text{EO}}\approx\,$20.000) \cite{gorecki1995physical}. It is worth mentioning that the experimental study also reported an Arrhenius behavior of the diffusion coefficients over a much larger temperature range \SI{333}{K}\,$\leq T\leq$\,\SI{400}{K}, and that the computational work furthermore demonstrated only a weak chain length dependence of $E_A$ for both ions (23\,$\leq N_{\text{EO}}\leq$\,450).
\FloatBarrier

\begin{figure}[!htb]
    \centering
    \begin{subfigure}[t]{0.9\linewidth}
        \includegraphics[width=\textwidth]{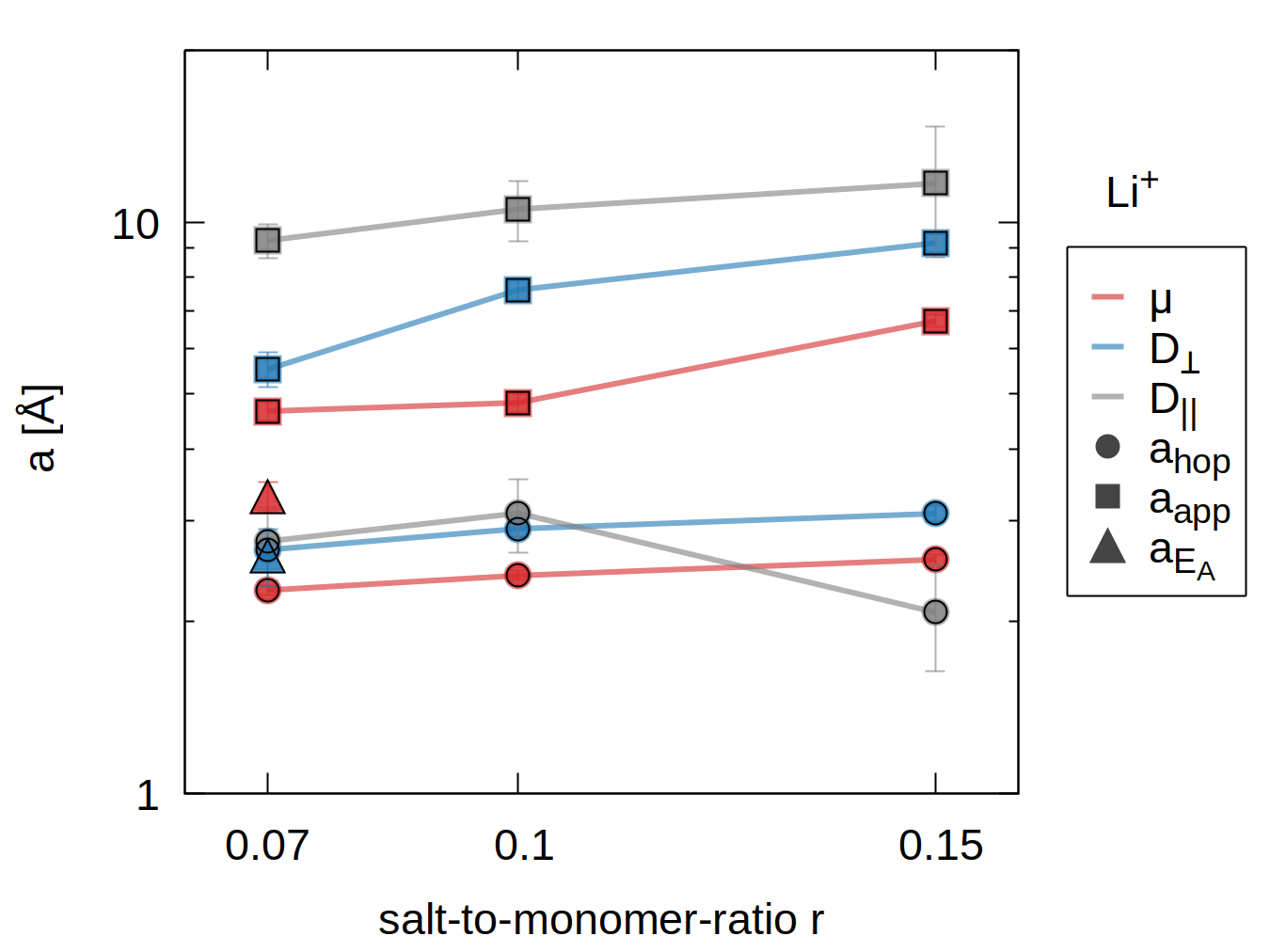}
        \caption{\Li hopping distances.}
        \label{fig:true_apparent_hopping_distances_lithium}
    \end{subfigure}

  \caption{Comparison of length scales $a_{\text{hop}}$, $a_{\text{app}}$ and $a_{\text{E}_\text{A}}$ derived from the field dependencies of \moby{}, \Dortho{} and \Dpara{} at various salt concentrations. The lines serve as a guide to the eye.} 
  \label{fig:hopping_distances_lithium_tfsi}
\end{figure}

In particular, we find that $E_A$ decreases linearly with \Ez{} in the high-field regime. From the theoretical analysis we expect $E_A(E_z)\,= V_0 - q\cdot a/2\cdot E_z$ where the value of $a$ was denoted \aea{}. As shown in Figure \ref{fig:hopping_distances_lithium_tfsi}, the \aea{} values agree well with the previously determined \ahop{}. 
We mention in passing that the plateau-like behavior for small \Ez{} should not be interpreted as statistical scattering. Rather this behavior is to be expected in the linear response regime of very low fields. 

Lastly, the decrease of $E_A$ with electric field as shown in Figure \ref{fig:E_A_function_of_Ez} provides an intuitive explanation of why simulations at even higher \Ez{}, \eg{}, 3\,V/nm, turn out to be unstable: due to vanishing barriers, the ion motion becomes unrestrained and results in a physical instability of the simulation box.

\section{Discussion and summary}

In this work, we studied the extent to which nonlinear ion dynamics in PEO/LiTFSI electrolytes under the application of an external electric field \Ez{} can be understood based on a simple hopping model. In this theoretical framework, a nonlinear dynamic enhancement can be trivially explained by the fact that the field lowers the activation barriers for ion transport by tilting the potential energy surface in field direction. 

In taking this approach, we have drawn inspiration from the research field of ion-conducting glasses where crucial properties of  dynamic nonlinearities could be rationalized by modeling single particle hopping on an energetically disordered lattice \cite{heuer2005nonlinear,heuer2014physical, mattner2014frequency}.
Naturally, this ansatz relies on the assumption that the complex ion motion can be effectively discretised into jumps between homogeneously distributed sites, which are an effective distance $a$ apart. Given the strong interaction of \Li{} with  ether oxygens \opeo{}, which guide the \Li{} motion in the polymer host, a structural estimator of $a$ was obtained from their typical next-neighbor distance \astruc{}$\,\approx\,$2\,\angstrom{}.

Under the assumption that that the spatially disordered polymer matrix, and hence the hypothetical hopping lattice, is not structurally aligned with the direction of \Ez{}, we generalized the one-dimensional hopping model to arbitrary jump angles. While seemingly counterintuitive, orthogonal dynamics benefits from the reduced barrier heights in the field direction due to the very fact that ion jumps occur at random orientation relative to \Ez{}.
Thus, we derived analytical expressions for the field dependence of ion mobility \moby{}(\Ez{}) as well as parallel \Dpara{}(\Ez{}) and orthogonal \Dortho{}(\Ez{}) diffusion coefficients. 
We then used these relations to extract two types of ion hopping distances from the simulation data covering a broad range of \Ez{}: (1) apparent hopping distances \aapp{} from the nonlinear dynamics in the low-field regime and (2) \ahop{} and \aea{} from analysis of the high-field regime, respectively. 
Figure \ref{fig:hopping_distances_lithium_tfsi} summarizes the results for increasing salt concentration in the polymer electrolyte.

The general attributes of the energy landscapes explored by \Li{} could be tentatively elucidated by means of \aapp{} analogous to previous theoretical modeling \cite{heuer2014physical}. We found \aapp{} to be significantly larger than \astruc{}, suggesting that the energetic disorder is dominated by low-energy sites, \eg{}, \Li{} captured by crown-ether arrangements of the PEO chains. 
The fact that the apparent hopping distances vary for the different transport observables suggests that the disorder of the PEL reflects to a differing degree the respective transport properties. Indeed, it is demonstrated in Appendix \ref{appendix:PEL_disorder_app_d_mu} for \moby{} and \Dpara{} for the example of the modeling framework of Ref. \cite{heuer2014physical} that the results do not need to be identical.

\begin{figure}[H]
    \centering
    \begin{subfigure}[t]{0.9\linewidth}
        \includegraphics[width=\textwidth]{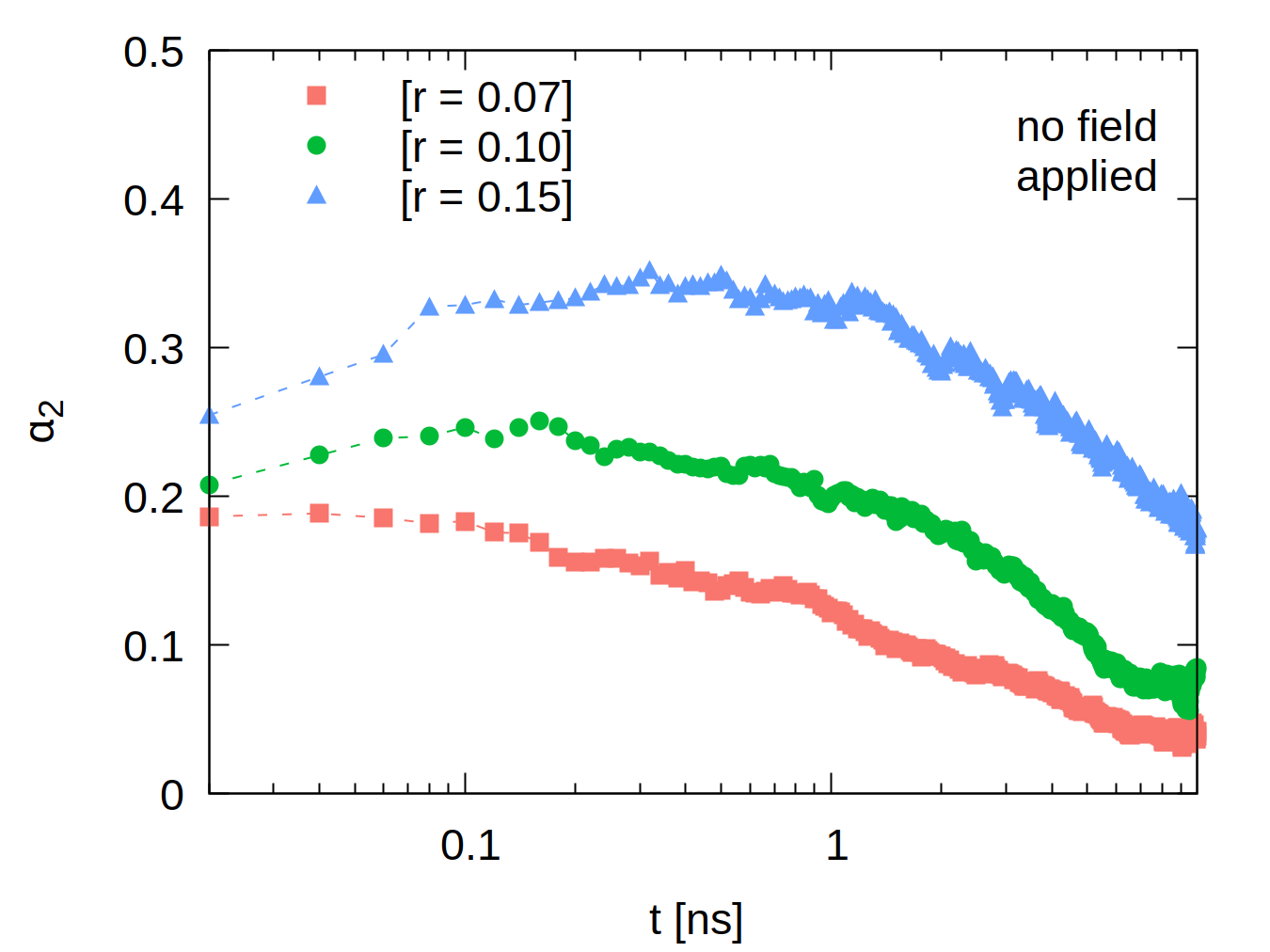}
    \end{subfigure}
  \caption{Non-Gaussian parameter $\alpha_2$ of the \Li{} dynamics when no field is applied for different values of the salt concentration.}
  \label{fig:A5_ngp_no_field}
\end{figure}

Interestingly, we observed that \aapp{} increases systematically for both ion species with increasing salt content $r$ which could indicate that the disorder of the PEL increases.
This can be directly corroborated by studying the non-Gaussian parameter 
\begin{equation}
    \alpha_2(t) = \frac{3 \langle \Delta r^4 (t)\rangle }{ 5\langle \Delta r^2 (t)\rangle^2}-1
\end{equation}
which may be considered as a measure of dynamic heterogeneities \cite{Shell_2005}. The results are shown in Fig. \ref{fig:A5_ngp_no_field}. Here one can clearly see a systematic increase with increasing salt concentration. The maximum value for the highest salt concentration is below 0.4. In contrast, for simulations of silicate systems the non-Gaussian parameter is at least one order of magnitude higher \cite{BALBUENA2014124}. This is consistent with the observation that \aapp{} for the lithium dynamics in inorganic ion conductors (approx. 4 nm) is much larger than the values found in  the present work. One possible contribution to the smallness of the non-Gaussian parameter for polymer electrolytes is the presence of fluctuating barriers that effectively reduce the dynamic heterogeneities.  Thus, neither our simulation results nor the general reasoning, based on the non-Gaussian parameter, is compatible with the large value  of \aapp{} of approx. 100 nm for the PEO/LiClO$_4$-system, reported in \cite{tajitsu1996linear}.

Building on the theoretical understanding of hopping dynamics in disordered energy landscapes \cite{mattner2014frequency,heuer2005nonlinear} according to which ion dynamics would be no longer trapped by the energetic disorder at high \Ez{}, we could identify \ahop{} and \aea{} with the true ion hopping distances. Indeed, the resulting values agree quite well with the structural estimator \astruc{}. The small but significant differences of the results from  \moby{}  and \Dortho{} may be related to the structural anisotropy of the polymer chains which increase with the size of the electric field.

In conclusion, our analysis showed that complex nonlinear ion dynamics in a soft matter electrolyte are well captured within the theoretical framework of a simple hopping picture in disordered energy landscapes.


\begin{acknowledgments}
We acknowledge the usage of the PALMA computer cluster at the University of Muenster.
\end{acknowledgments}

\appendix




\section{Taylor expansions of Equations \ref{eq:exact_solution_mu_u}, \ref{eq:exact_solution_d_para_u} and \ref{eq:exact_solution_d_ortho_u} for weak \Ez{}: $\mu/D_{\parallel}/D_{\perp} = \lambda_0 + \lambda_2\cdot E^2$}
\label{appendix:Taylor_expansion}

With $u\,=\,q\cdot\dfrac{a}{2}\cdot\beta\cdot E_z\quad = \quad \alpha \cdot E_z$ \newline
\newline

$\Big(\Lim{E_z \to 0}\mu(E_z)$ starting from Eq. \ref{eq:exact_solution_mu_u}$\Big)$:

\begingroup
    \fontsize{8pt}{12pt}\selectfont
\begin{equation}
  \begin{alignedat}{3}
     \mu(E_z) =& &&a\Gamma_0&&\Big[ \dfrac{\cosh(\alpha E_z)}{\alpha E_z^2}  +  \dfrac{\sinh(\alpha E_z)}{\alpha^2 E_z^3}  \Big] \\
     \stackrel{\text {Taylor}}{=}&  &&a\Gamma_0&&\Big[ \dfrac{\alpha}{3}  +  \dfrac{\alpha^3E_z^2}{30} \quad + \mathcal{O}(E_z^4) \Big] \\
     \approx& && &&\Big[\dfrac{\Gamma_0\cdot q\cdot a^2\cdot\beta }{6}\Big] + \Big[\dfrac{a\Gamma_0 }{240}(q\cdot a\cdot\beta)^3\Big]\cdot E_z^2 \\
     =& && &&\lambda_0 + \lambda_2\cdot E_z^2
    \end{alignedat}
     \label{eq:A5_taylor_mu}
\end{equation}
  
\endgroup

$\Big(\Lim{E_z \to 0}D_\parallel(E_z)$ starting from Eq. \ref{eq:exact_solution_d_para_u}$\Big)$:

\begingroup
    \fontsize{8pt}{12pt}\selectfont
\begin{equation}
  \begin{alignedat}{3}
     D_{\parallel}(E_z) =& &&\dfrac{a^2\Gamma_0}{2}&&\Big[ \dfrac{(\alpha^2E_z^2 +2) \sinh(\alpha E_z)}{(\alpha E_z)^3} - \dfrac{2\cosh(\alpha E_z)}{(\alpha E_z)^2} \Big] \\
     \stackrel{\text {Taylor}}{=}&  &&\dfrac{a^2\Gamma_0}{2}&&\Big[ \dfrac{1}{3}  +  \dfrac{\alpha^2E_z^2}{10} \quad + \mathcal{O}(E_z^4) \Big] \\
     \approx& && &&\Big[\dfrac{\Gamma_0\cdot\cdot a^2 }{6}\Big] + \Big[\dfrac{a^2\Gamma_0 }{80}(q\cdot a\cdot\beta)^2\Big]\cdot E_z^2 \\
     &= && &&\lambda_0 + \lambda_2\cdot E_z^2
    \end{alignedat}
     \label{eq:A5_taylor_d_para}
\end{equation}
\endgroup

Please note that the linear-response mobility $\mu(E_z=0)$ and the unperturbed self-diffusion coefficient $D_\parallel(E_z=0)$ are related through the Einstein-relation $D_\parallel = \mu k_B T\,/\,q$, which is also fulfilled after the angular-average procedure when comparing $\lambda_{0,D_\parallel} = \lambda_{0,\mu}\,/(q\cdot \beta)$.

$\Big(\Lim{E_z \to 0}D_\perp(E_z)$ starting from Eq. \ref{eq:exact_solution_d_ortho_u}$\Big)$:

\begingroup
    \fontsize{8pt}{12pt}\selectfont
\begin{equation}
  \begin{alignedat}{3}
     D_{\perp}(E_z) =& &&\dfrac{a^2\Gamma_0}{2}&&\Big[ \dfrac{\cosh(\alpha E_z)}{(\alpha E_z)^2} - \dfrac{\sinh(\alpha E_z)}{(\alpha E_z)^3} \Big] \\
     \stackrel{\text {Taylor}}{=}&  &&\dfrac{a^2\Gamma_0}{2}&&\Big[ \dfrac{1}{3}  +  \dfrac{\alpha^2E_z^2}{30} \quad + \mathcal{O}(E_z^4) \Big] \\
     \approx& && &&\Big[\dfrac{\Gamma_0\cdot a^2 }{6}\Big] + \Big[\dfrac{a^2\Gamma_0 }{240}(q\cdot a\cdot\beta)^2\Big]\cdot E_z^2 \\
     =& && &&\lambda_0 + \lambda_2\cdot E_z^2
    \end{alignedat}
     \label{eq:A5_taylor_d_ortho}
\end{equation}
\endgroup

\section{Impact of disorder on $a_{\text{app}}$ of current and diffusivity in limit of weak \Ez{}} \label{appendix:PEL_disorder_app_d_mu}

The impact of energetic disorder on \aapp{} and the underlying mechanisms of nonlinear ion transport have been explored in Ref. \cite{heuer2014physical} for a $d-$dimensional, bimodal energy landscape for which analytical dependencies of \aapp{} on the parameters quantifying the disorder could be derived.
The lattice sites are a distance $a$ apart and can assume randomly chosen energies of $\pm\Delta U/2$ so that the energetic disorder can be tuned by means of (1) the energy difference $\Delta U$, which is described by $Z=\exp(\beta\Delta U/2)$, and (2) the relative decomposition of all sites into low- and high-energy levels, which is indicated by the fraction $\alpha$ of low-energy sites. The number of hopping pathways orthogonal to the electric field direction is given by $\Theta = 2\cdot(d-1)$.

In  case of a small number of low-energy sites (formally expressed as $\Theta\ll Z$ and $\alpha\cdot\Theta\ll 1$) one obtains the simple physical picture that the fraction of particles in the high-energy sites is increased by a factor 
\begin{equation}
p_+ = 1 + k(\Theta, Z,\alpha)\cdot\sinh(u)^2
\end{equation}
as compared to the equilibrium population (see Equation (13) in Ref. \cite{heuer2014physical}) where the positive value $k$ can be expressed analytically. Furthermore, the particles in the low-energy sites can be considered as trapped independent of the applied electric field.

For the hopping dynamics in a regular energy landscape with a field along the hopping direction one has (see Eq. \ref{eq:sinh_relation_j_1D_basis}) $j \propto \sinh(u)$. In analogy one obtains $D_\parallel(u) \propto \cosh(u)^2$. This can be easily seen from the derivation in Eq.\ref{eq:exact_solution_d_para_u} when instead of integrating over all $\Theta$ one just adds over $\Theta = 0$ and $\Theta = \pi$, see also \cite{kunow2006nonlinear}. Furthermore, both values are proportional to the number of mobile particles, varying with the applied field. Thus, we finally get 

\begin{equation}
  \begin{alignedat}{1} 
    j(u) &=  (1 + k \sinh^2(u)  )\sinh(u) \\
    &\approx u + \frac{u^3}{6} ( 1+6k  )
    \end{alignedat}
\end{equation}
and 
\begin{equation}
  \begin{alignedat}{1} 
    D_\parallel(u) &=   (1 + k \sinh^2(u)  )\cosh(u) \\
     &\approx 1 + \frac{u^2}{2}  ( 1+2k  ).
    \end{alignedat}
\end{equation}

From this representation the impact of the disorder (via $k$) on $a_{\text{app}}/a$ can be directly identified from the factors $(1+6k)^{0.5}$ and $(1 + 2k)^{0.5}$, respectively. Interestingly, the increase of $a_{\text{app}}$ is not identical for both observables as also found in this work. Note, however, that for the lattice model the increase of $a_{\text{app}}$ is more pronounced for $j$ as compared to $D_\parallel$ in contrast to our simulation results. However, this discrepancy should not be interpreted too strongly since the type of disorder in the model as studied in \cite{heuer2014physical} has little similarity with the potential energy landscape as experienced by the ions in a polymer electrolyte.

\section{High-field regime for d-dimensional hopping model in disordered energy landscape} \label{appendix:hopping_model_analytical}

Using the notation from the main text and taking the results from Ref. \cite{mattner2014frequency}, the resulting stationary current upon application of a constant electric field reads (see Equations 11 and 28 in Ref.\cite{mattner2014frequency}) 
\begin{widetext}
\begin{equation}
    \begin{alignedat}{2}
&j(u) =  &&-\Gamma_2e^{-u} +\Gamma_2\Gamma_3 e^{u} + y_1(u)\,\cdot\,\\ 
         &      && \left[ \left(\Gamma_1-\Gamma_2\Gamma_3\right) e^{u} + \left(\Gamma_2 - \Gamma_1\Gamma_3\right)e^{-u} \right] \\
    \text{with}\qquad 
     & {y}_1(u) =  &&\Gamma_2 \cdot \frac{  (e^{-u} + d -1)+ \Gamma_3 \cdot(e^{u} + d -1)}{(\Gamma_1+\Gamma_2\Gamma_3)\cdot(e^{u} + d -1) + (\Gamma_2+\Gamma_1\Gamma_3)\cdot (e^{-u} + d -1)}
    \end{alignedat}
\end{equation}
\end{widetext}
    Here, the $\Gamma_i=\exp{-\beta V_i}$ parameters are associated with the jumping rates between the disordered potential barriers which can be expressed through $V_i$. $y_1$ denotes the population of site 1, see Figure \ref{fig:A5_scheme_adapted_from_Mattner2014}. The field dependency is contained in $u =  \beta q a  E_z/2$. In the high-field regime, the behavior of $j(u)$ is dominated by the exponential term  $\exp(u)$.
    
    Straightforward calculation yields in this limit
    \begin{equation}
        j(u) = 2 \frac{\Gamma_1 \Gamma_2 \Gamma_3}{\Gamma_1 + \Gamma_2 \Gamma_3 }\exp(u) + f(\Gamma_1, \Gamma_2,\Gamma_3,d) + O(e^{-u})
    \end{equation}
     with some function $f(.)$. Its specific form is not relevant for the present analysis.

\bibliography{myBibliography}

\providecommand{\noopsort}[1]{}\providecommand{\singleletter}[1]{#1}%
\begin{thebibliography}{41}%
\makeatletter
\providecommand \@ifxundefined [1]{%
 \@ifx{#1\undefined}
}%
\providecommand \@ifnum [1]{%
 \ifnum #1\expandafter \@firstoftwo
 \else \expandafter \@secondoftwo
 \fi
}%
\providecommand \@ifx [1]{%
 \ifx #1\expandafter \@firstoftwo
 \else \expandafter \@secondoftwo
 \fi
}%
\providecommand \natexlab [1]{#1}%
\providecommand \enquote  [1]{``#1''}%
\providecommand \bibnamefont  [1]{#1}%
\providecommand \bibfnamefont [1]{#1}%
\providecommand \citenamefont [1]{#1}%
\providecommand \href@noop [0]{\@secondoftwo}%
\providecommand \href [0]{\begingroup \@sanitize@url \@href}%
\providecommand \@href[1]{\@@startlink{#1}\@@href}%
\providecommand \@@href[1]{\endgroup#1\@@endlink}%
\providecommand \@sanitize@url [0]{\catcode `\\12\catcode `\$12\catcode
  `\&12\catcode `\#12\catcode `\^12\catcode `\_12\catcode `\%12\relax}%
\providecommand \@@startlink[1]{}%
\providecommand \@@endlink[0]{}%
\providecommand \url  [0]{\begingroup\@sanitize@url \@url }%
\providecommand \@url [1]{\endgroup\@href {#1}{\urlprefix }}%
\providecommand \urlprefix  [0]{URL }%
\providecommand \Eprint [0]{\href }%
\providecommand \doibase [0]{http://dx.doi.org/}%
\providecommand \selectlanguage [0]{\@gobble}%
\providecommand \bibinfo  [0]{\@secondoftwo}%
\providecommand \bibfield  [0]{\@secondoftwo}%
\providecommand \translation [1]{[#1]}%
\providecommand \BibitemOpen [0]{}%
\providecommand \bibitemStop [0]{}%
\providecommand \bibitemNoStop [0]{.\EOS\space}%
\providecommand \EOS [0]{\spacefactor3000\relax}%
\providecommand \BibitemShut  [1]{\csname bibitem#1\endcsname}%
\let\auto@bib@innerbib\@empty
\bibitem [{\citenamefont {Wettstein}(2022)}]{wettstein2022dissertation}%
  \BibitemOpen
  \bibfield  {author} {\bibinfo {author} {\bibfnamefont {A.}~\bibnamefont
  {Wettstein}},\ }\bibfield  {title} {\enquote {\bibinfo {title} {Towards a
  mechanistic understanding of lithium transport in electrolytes: Insights from
  molecular dynamics simulations},}\ }\href@noop {} {\bibfield  {journal}
  {\bibinfo  {journal} {Westf{\"a}lische Wilhelms-Universit{\"a}t M{\"u}nster,
  Dissertation}\ } (\bibinfo {year} {2022})}\BibitemShut {NoStop}%
\bibitem [{\citenamefont {Ferrari}\ \emph {et~al.}(2015)\citenamefont
  {Ferrari}, \citenamefont {Loveridge}, \citenamefont {Beattie}, \citenamefont
  {Jahn}, \citenamefont {Dashwood},\ and\ \citenamefont
  {Bhagat}}]{ferrari2015latest}%
  \BibitemOpen
  \bibfield  {author} {\bibinfo {author} {\bibfnamefont {S.}~\bibnamefont
  {Ferrari}}, \bibinfo {author} {\bibfnamefont {M.}~\bibnamefont {Loveridge}},
  \bibinfo {author} {\bibfnamefont {S.~D.}\ \bibnamefont {Beattie}}, \bibinfo
  {author} {\bibfnamefont {M.}~\bibnamefont {Jahn}}, \bibinfo {author}
  {\bibfnamefont {R.~J.}\ \bibnamefont {Dashwood}}, \ and\ \bibinfo {author}
  {\bibfnamefont {R.}~\bibnamefont {Bhagat}},\ }\bibfield  {title} {\enquote
  {\bibinfo {title} {Latest advances in the manufacturing of 3d rechargeable
  lithium microbatteries},}\ }\href@noop {} {\bibfield  {journal} {\bibinfo
  {journal} {Journal of Power Sources}\ }\textbf {\bibinfo {volume} {286}},\
  \bibinfo {pages} {25--46} (\bibinfo {year} {2015})}\BibitemShut {NoStop}%
\bibitem [{\citenamefont {Ni}\ \emph {et~al.}(2020)\citenamefont {Ni},
  \citenamefont {Dai}, \citenamefont {Yuan}, \citenamefont {Li},\ and\
  \citenamefont {Lu}}]{ni2020three}%
  \BibitemOpen
  \bibfield  {author} {\bibinfo {author} {\bibfnamefont {J.}~\bibnamefont
  {Ni}}, \bibinfo {author} {\bibfnamefont {A.}~\bibnamefont {Dai}}, \bibinfo
  {author} {\bibfnamefont {Y.}~\bibnamefont {Yuan}}, \bibinfo {author}
  {\bibfnamefont {L.}~\bibnamefont {Li}}, \ and\ \bibinfo {author}
  {\bibfnamefont {J.}~\bibnamefont {Lu}},\ }\bibfield  {title} {\enquote
  {\bibinfo {title} {Three-dimensional microbatteries beyond lithium ion},}\
  }\href@noop {} {\bibfield  {journal} {\bibinfo  {journal} {Matter}\ }\textbf
  {\bibinfo {volume} {2}},\ \bibinfo {pages} {1366--1376} (\bibinfo {year}
  {2020})}\BibitemShut {NoStop}%
\bibitem [{\citenamefont {Jetybayeva}\ \emph {et~al.}(2021)\citenamefont
  {Jetybayeva}, \citenamefont {Uzakbaiuly}, \citenamefont {Mukanova},
  \citenamefont {Myung},\ and\ \citenamefont {Bakenov}}]{jetybayeva2021recent}%
  \BibitemOpen
  \bibfield  {author} {\bibinfo {author} {\bibfnamefont {A.}~\bibnamefont
  {Jetybayeva}}, \bibinfo {author} {\bibfnamefont {B.}~\bibnamefont
  {Uzakbaiuly}}, \bibinfo {author} {\bibfnamefont {A.}~\bibnamefont
  {Mukanova}}, \bibinfo {author} {\bibfnamefont {S.-T.}\ \bibnamefont {Myung}},
  \ and\ \bibinfo {author} {\bibfnamefont {Z.}~\bibnamefont {Bakenov}},\
  }\bibfield  {title} {\enquote {\bibinfo {title} {Recent advancements in solid
  electrolytes integrated into all-solid-state 2d and 3d lithium-ion
  microbatteries},}\ }\href@noop {} {\bibfield  {journal} {\bibinfo  {journal}
  {Journal of Materials Chemistry A}\ }\textbf {\bibinfo {volume} {9}},\
  \bibinfo {pages} {15140--15178} (\bibinfo {year} {2021})}\BibitemShut
  {NoStop}%
\bibitem [{\citenamefont {Roling}\ \emph {et~al.}(2008)\citenamefont {Roling},
  \citenamefont {Murugavel}, \citenamefont {Heuer}, \citenamefont
  {L{\"u}hning}, \citenamefont {Friedrich},\ and\ \citenamefont
  {R{\"o}thel}}]{roling2008field}%
  \BibitemOpen
  \bibfield  {author} {\bibinfo {author} {\bibfnamefont {B.}~\bibnamefont
  {Roling}}, \bibinfo {author} {\bibfnamefont {S.}~\bibnamefont {Murugavel}},
  \bibinfo {author} {\bibfnamefont {A.}~\bibnamefont {Heuer}}, \bibinfo
  {author} {\bibfnamefont {L.}~\bibnamefont {L{\"u}hning}}, \bibinfo {author}
  {\bibfnamefont {R.}~\bibnamefont {Friedrich}}, \ and\ \bibinfo {author}
  {\bibfnamefont {S.}~\bibnamefont {R{\"o}thel}},\ }\bibfield  {title}
  {\enquote {\bibinfo {title} {Field-dependent ion transport in disordered
  solid electrolytes},}\ }\href@noop {} {\bibfield  {journal} {\bibinfo
  {journal} {Physical Chemistry Chemical Physics}\ }\textbf {\bibinfo {volume}
  {10}},\ \bibinfo {pages} {4211--4226} (\bibinfo {year} {2008})}\BibitemShut
  {NoStop}%
\bibitem [{\citenamefont {Ponce}, \citenamefont {Galvez-Aranda},\ and\
  \citenamefont {Seminario}(2021)}]{ponce2021analysis}%
  \BibitemOpen
  \bibfield  {author} {\bibinfo {author} {\bibfnamefont {V.}~\bibnamefont
  {Ponce}}, \bibinfo {author} {\bibfnamefont {D.~E.}\ \bibnamefont
  {Galvez-Aranda}}, \ and\ \bibinfo {author} {\bibfnamefont {J.~M.}\
  \bibnamefont {Seminario}},\ }\bibfield  {title} {\enquote {\bibinfo {title}
  {Analysis of an all-solid state nanobattery using molecular dynamics
  simulations under an external electric field},}\ }\href@noop {} {\bibfield
  {journal} {\bibinfo  {journal} {Physical Chemistry Chemical Physics}\
  }\textbf {\bibinfo {volume} {23}},\ \bibinfo {pages} {597--606} (\bibinfo
  {year} {2021})}\BibitemShut {NoStop}%
\bibitem [{\citenamefont {Matse}, \citenamefont {Berg},\ and\ \citenamefont
  {Eikerling}(2020)}]{Matse2020}%
  \BibitemOpen
  \bibfield  {author} {\bibinfo {author} {\bibfnamefont {M.}~\bibnamefont
  {Matse}}, \bibinfo {author} {\bibfnamefont {P.}~\bibnamefont {Berg}}, \ and\
  \bibinfo {author} {\bibfnamefont {M.}~\bibnamefont {Eikerling}},\ }\bibfield
  {title} {\enquote {\bibinfo {title} {{Asymmetric double-layer charging in a
  cylindrical nanopore under closed confinement }},}\ }\href {\doibase
  10.1063/1.5139541} {\bibfield  {journal} {\bibinfo  {journal} {The Journal of
  Chemical Physics}\ }\textbf {\bibinfo {volume} {152}} (\bibinfo {year}
  {2020}),\ 10.1063/1.5139541}\BibitemShut {NoStop}%
\bibitem [{\citenamefont {Jorn}\ \emph {et~al.}(2013)\citenamefont {Jorn},
  \citenamefont {Kumar}, \citenamefont {Abraham},\ and\ \citenamefont
  {Voth}}]{Jorn2013}%
  \BibitemOpen
  \bibfield  {author} {\bibinfo {author} {\bibfnamefont {R.}~\bibnamefont
  {Jorn}}, \bibinfo {author} {\bibfnamefont {R.}~\bibnamefont {Kumar}},
  \bibinfo {author} {\bibfnamefont {D.}~\bibnamefont {Abraham}}, \ and\
  \bibinfo {author} {\bibfnamefont {G.}~\bibnamefont {Voth}},\ }\bibfield
  {title} {\enquote {\bibinfo {title} {{Atomistic Modeling of the
  Electrode-Electrolyte Interface in Li-Ion Energy Storage Systems: Electrolyte
  Structuring}},}\ }\href {\doibase 10.1021/jp3102282} {\bibfield  {journal}
  {\bibinfo  {journal} {The Journal of Physical Chemistry C}\ }\textbf
  {\bibinfo {volume} {117}},\ \bibinfo {pages} {3747--3761} (\bibinfo {year}
  {2013})}\BibitemShut {NoStop}%
\bibitem [{\citenamefont {Stuve}(2012)}]{Stuve2011}%
  \BibitemOpen
  \bibfield  {author} {\bibinfo {author} {\bibfnamefont {E.}~\bibnamefont
  {Stuve}},\ }\bibfield  {title} {\enquote {\bibinfo {title} {{Ionization of
  water in interfacial electric fields: An electrichemical view}},}\ }\href
  {\doibase 10.1016/j.cplett.2011.09.040} {\bibfield  {journal} {\bibinfo
  {journal} {Chemical Physics Letters}\ }\textbf {\bibinfo {volume}
  {519-520}},\ \bibinfo {pages} {1--17} (\bibinfo {year} {2012})}\BibitemShut
  {NoStop}%
\bibitem [{\citenamefont {Lepley}\ and\ \citenamefont
  {Holzwarth}(2015)}]{lepley2015modeling}%
  \BibitemOpen
  \bibfield  {author} {\bibinfo {author} {\bibfnamefont {N.}~\bibnamefont
  {Lepley}}\ and\ \bibinfo {author} {\bibfnamefont {N.}~\bibnamefont
  {Holzwarth}},\ }\bibfield  {title} {\enquote {\bibinfo {title} {Modeling
  interfaces between solids: Application to li battery materials},}\
  }\href@noop {} {\bibfield  {journal} {\bibinfo  {journal} {Physical Review
  B}\ }\textbf {\bibinfo {volume} {92}},\ \bibinfo {pages} {214201} (\bibinfo
  {year} {2015})}\BibitemShut {NoStop}%
\bibitem [{\citenamefont {Kunow}\ and\ \citenamefont
  {Heuer}(2006)}]{kunow2006nonlinear}%
  \BibitemOpen
  \bibfield  {author} {\bibinfo {author} {\bibfnamefont {M.}~\bibnamefont
  {Kunow}}\ and\ \bibinfo {author} {\bibfnamefont {A.}~\bibnamefont {Heuer}},\
  }\bibfield  {title} {\enquote {\bibinfo {title} {Nonlinear ionic conductivity
  of lithium silicate glass studied via molecular dynamics simulations},}\
  }\href@noop {} {\bibfield  {journal} {\bibinfo  {journal} {The Journal of
  chemical physics}\ }\textbf {\bibinfo {volume} {124}},\ \bibinfo {pages}
  {214703} (\bibinfo {year} {2006})}\BibitemShut {NoStop}%
\bibitem [{\citenamefont {Heuer}, \citenamefont {Murugavel},\ and\
  \citenamefont {Roling}(2005)}]{heuer2005nonlinear}%
  \BibitemOpen
  \bibfield  {author} {\bibinfo {author} {\bibfnamefont {A.}~\bibnamefont
  {Heuer}}, \bibinfo {author} {\bibfnamefont {S.}~\bibnamefont {Murugavel}}, \
  and\ \bibinfo {author} {\bibfnamefont {B.}~\bibnamefont {Roling}},\
  }\bibfield  {title} {\enquote {\bibinfo {title} {Nonlinear ionic conductivity
  of thin solid electrolyte samples: Comparison between theory and
  experiment},}\ }\href@noop {} {\bibfield  {journal} {\bibinfo  {journal}
  {Physical Review B}\ }\textbf {\bibinfo {volume} {72}},\ \bibinfo {pages}
  {174304} (\bibinfo {year} {2005})}\BibitemShut {NoStop}%
\bibitem [{\citenamefont {Heuer}\ and\ \citenamefont
  {L{\"u}hning}(2014)}]{heuer2014physical}%
  \BibitemOpen
  \bibfield  {author} {\bibinfo {author} {\bibfnamefont {A.}~\bibnamefont
  {Heuer}}\ and\ \bibinfo {author} {\bibfnamefont {L.}~\bibnamefont
  {L{\"u}hning}},\ }\bibfield  {title} {\enquote {\bibinfo {title} {Physical
  mechanisms of nonlinear conductivity: A model analysis},}\ }\href@noop {}
  {\bibfield  {journal} {\bibinfo  {journal} {The Journal of Chemical Physics}\
  }\textbf {\bibinfo {volume} {140}},\ \bibinfo {pages} {094508} (\bibinfo
  {year} {2014})}\BibitemShut {NoStop}%
\bibitem [{\citenamefont {Brooks}\ \emph {et~al.}(2018)\citenamefont {Brooks},
  \citenamefont {Merinov}, \citenamefont {Goddard~III}, \citenamefont
  {Kozinsky},\ and\ \citenamefont {Mailoa}}]{brooks2018atomistic}%
  \BibitemOpen
  \bibfield  {author} {\bibinfo {author} {\bibfnamefont {D.~J.}\ \bibnamefont
  {Brooks}}, \bibinfo {author} {\bibfnamefont {B.~V.}\ \bibnamefont {Merinov}},
  \bibinfo {author} {\bibfnamefont {W.~A.}\ \bibnamefont {Goddard~III}},
  \bibinfo {author} {\bibfnamefont {B.}~\bibnamefont {Kozinsky}}, \ and\
  \bibinfo {author} {\bibfnamefont {J.}~\bibnamefont {Mailoa}},\ }\bibfield
  {title} {\enquote {\bibinfo {title} {Atomistic description of ionic diffusion
  in peo--litfsi: Effect of temperature, molecular weight, and ionic
  concentration},}\ }\href@noop {} {\bibfield  {journal} {\bibinfo  {journal}
  {Macromolecules}\ }\textbf {\bibinfo {volume} {51}},\ \bibinfo {pages}
  {8987--8995} (\bibinfo {year} {2018})}\BibitemShut {NoStop}%
\bibitem [{\citenamefont {Maitra}\ and\ \citenamefont
  {Heuer}(2007)}]{maitra2007cation}%
  \BibitemOpen
  \bibfield  {author} {\bibinfo {author} {\bibfnamefont {A.}~\bibnamefont
  {Maitra}}\ and\ \bibinfo {author} {\bibfnamefont {A.}~\bibnamefont {Heuer}},\
  }\bibfield  {title} {\enquote {\bibinfo {title} {Cation transport in polymer
  electrolytes: a microscopic approach},}\ }\href@noop {} {\bibfield  {journal}
  {\bibinfo  {journal} {Physical review letters}\ }\textbf {\bibinfo {volume}
  {98}},\ \bibinfo {pages} {227802} (\bibinfo {year} {2007})}\BibitemShut
  {NoStop}%
\bibitem [{\citenamefont {Diddens}, \citenamefont {Heuer},\ and\ \citenamefont
  {Borodin}(2010)}]{diddens2010understanding}%
  \BibitemOpen
  \bibfield  {author} {\bibinfo {author} {\bibfnamefont {D.}~\bibnamefont
  {Diddens}}, \bibinfo {author} {\bibfnamefont {A.}~\bibnamefont {Heuer}}, \
  and\ \bibinfo {author} {\bibfnamefont {O.}~\bibnamefont {Borodin}},\
  }\bibfield  {title} {\enquote {\bibinfo {title} {Understanding the lithium
  transport within a rouse-based model for a peo/litfsi polymer electrolyte},}\
  }\href@noop {} {\bibfield  {journal} {\bibinfo  {journal} {Macromolecules}\
  }\textbf {\bibinfo {volume} {43}},\ \bibinfo {pages} {2028--2036} (\bibinfo
  {year} {2010})}\BibitemShut {NoStop}%
\bibitem [{\citenamefont {Tajitsu}(1996)}]{tajitsu1996linear}%
  \BibitemOpen
  \bibfield  {author} {\bibinfo {author} {\bibfnamefont {Y.}~\bibnamefont
  {Tajitsu}},\ }\bibfield  {title} {\enquote {\bibinfo {title} {Linear and
  non-linear conductive spectra of polyethylene oxide/salt complex},}\
  }\href@noop {} {\bibfield  {journal} {\bibinfo  {journal} {Journal of
  materials science}\ }\textbf {\bibinfo {volume} {31}},\ \bibinfo {pages}
  {2081--2089} (\bibinfo {year} {1996})}\BibitemShut {NoStop}%
\bibitem [{\citenamefont {Rosenwinkel}\ and\ \citenamefont
  {Sch{\"{o}}nhoff}(2019)}]{Rosenwinkel2019}%
  \BibitemOpen
  \bibfield  {author} {\bibinfo {author} {\bibfnamefont {M.~P.}\ \bibnamefont
  {Rosenwinkel}}\ and\ \bibinfo {author} {\bibfnamefont {M.}~\bibnamefont
  {Sch{\"{o}}nhoff}},\ }\bibfield  {title} {\enquote {\bibinfo {title}
  {{Lithium Transference Numbers in PEO/LiTFSA Electrolytes Determined by
  Electrophoretic NMR}},}\ }\href {\doibase 10.1149/2.0831910jes} {\bibfield
  {journal} {\bibinfo  {journal} {Journal of The Electrochemical Society}\
  }\textbf {\bibinfo {volume} {166}},\ \bibinfo {pages} {A1977--A1983}
  (\bibinfo {year} {2019})}\BibitemShut {NoStop}%
\bibitem [{\citenamefont {Wettstein}, \citenamefont {Diddens},\ and\
  \citenamefont {Heuer}(2021)}]{wettstein2021polymer}%
  \BibitemOpen
  \bibfield  {author} {\bibinfo {author} {\bibfnamefont {A.}~\bibnamefont
  {Wettstein}}, \bibinfo {author} {\bibfnamefont {D.}~\bibnamefont {Diddens}},
  \ and\ \bibinfo {author} {\bibfnamefont {A.}~\bibnamefont {Heuer}},\
  }\bibfield  {title} {\enquote {\bibinfo {title} {Polymer electrolytes in
  strong external electric fields: Modification of structure and dynamics},}\
  }\href@noop {} {\bibfield  {journal} {\bibinfo  {journal} {Macromolecules}\
  }\textbf {\bibinfo {volume} {54}},\ \bibinfo {pages} {2256--2265} (\bibinfo
  {year} {2021})}\BibitemShut {NoStop}%
\bibitem [{\citenamefont {Mattner}, \citenamefont {Roling},\ and\ \citenamefont
  {Heuer}(2014)}]{mattner2014frequency}%
  \BibitemOpen
  \bibfield  {author} {\bibinfo {author} {\bibfnamefont {C.}~\bibnamefont
  {Mattner}}, \bibinfo {author} {\bibfnamefont {B.}~\bibnamefont {Roling}}, \
  and\ \bibinfo {author} {\bibfnamefont {A.}~\bibnamefont {Heuer}},\ }\bibfield
   {title} {\enquote {\bibinfo {title} {The frequency dependence of nonlinear
  conductivity in non-homogeneous systems: An analytically solvable model},}\
  }\href@noop {} {\bibfield  {journal} {\bibinfo  {journal} {Solid State
  Ionics}\ }\textbf {\bibinfo {volume} {261}},\ \bibinfo {pages} {28--35}
  (\bibinfo {year} {2014})}\BibitemShut {NoStop}%
\bibitem [{\citenamefont {Barton}(1996)}]{barton1996electric}%
  \BibitemOpen
  \bibfield  {author} {\bibinfo {author} {\bibfnamefont {J.}~\bibnamefont
  {Barton}},\ }\bibfield  {title} {\enquote {\bibinfo {title} {Electric
  conduction of glasses at intermediate field strengths},}\ }\href@noop {}
  {\bibfield  {journal} {\bibinfo  {journal} {Journal of non-crystalline
  solids}\ }\textbf {\bibinfo {volume} {203}},\ \bibinfo {pages} {280--285}
  (\bibinfo {year} {1996})}\BibitemShut {NoStop}%
\bibitem [{\citenamefont {Isard}(1996)}]{isard1996high}%
  \BibitemOpen
  \bibfield  {author} {\bibinfo {author} {\bibfnamefont {J.}~\bibnamefont
  {Isard}},\ }\bibfield  {title} {\enquote {\bibinfo {title} {High field
  conduction in ionic glasses—effect of a distribution of activation
  energies},}\ }\href@noop {} {\bibfield  {journal} {\bibinfo  {journal}
  {Journal of non-crystalline solids}\ }\textbf {\bibinfo {volume} {202}},\
  \bibinfo {pages} {137--144} (\bibinfo {year} {1996})}\BibitemShut {NoStop}%
\bibitem [{\citenamefont {Banhatti}\ and\ \citenamefont
  {Heuer}(2001)}]{banhatti2001structure}%
  \BibitemOpen
  \bibfield  {author} {\bibinfo {author} {\bibfnamefont {R.~D.}\ \bibnamefont
  {Banhatti}}\ and\ \bibinfo {author} {\bibfnamefont {A.}~\bibnamefont
  {Heuer}},\ }\bibfield  {title} {\enquote {\bibinfo {title} {Structure and
  dynamics of lithium silicate melts: molecular dynamics simulations},}\
  }\href@noop {} {\bibfield  {journal} {\bibinfo  {journal} {Physical Chemistry
  Chemical Physics}\ }\textbf {\bibinfo {volume} {3}},\ \bibinfo {pages}
  {5104--5108} (\bibinfo {year} {2001})}\BibitemShut {NoStop}%
\bibitem [{\citenamefont {Roling}(2002)}]{roling2002hopping}%
  \BibitemOpen
  \bibfield  {author} {\bibinfo {author} {\bibfnamefont {B.}~\bibnamefont
  {Roling}},\ }\bibfield  {title} {\enquote {\bibinfo {title} {Hopping dynamics
  of ions and polarons in disordered materials: On the potential of nonlinear
  conductivity spectroscopy},}\ }\href@noop {} {\bibfield  {journal} {\bibinfo
  {journal} {The Journal of chemical physics}\ }\textbf {\bibinfo {volume}
  {117}},\ \bibinfo {pages} {1320--1327} (\bibinfo {year} {2002})}\BibitemShut
  {NoStop}%
\bibitem [{\citenamefont {R{\"o}thel}\ \emph {et~al.}(2010)\citenamefont
  {R{\"o}thel}, \citenamefont {Friedrich}, \citenamefont {L{\"u}hning},\ and\
  \citenamefont {Heuer}}]{rothel2010theoretical}%
  \BibitemOpen
  \bibfield  {author} {\bibinfo {author} {\bibfnamefont {S.}~\bibnamefont
  {R{\"o}thel}}, \bibinfo {author} {\bibfnamefont {R.}~\bibnamefont
  {Friedrich}}, \bibinfo {author} {\bibfnamefont {L.}~\bibnamefont
  {L{\"u}hning}}, \ and\ \bibinfo {author} {\bibfnamefont {A.}~\bibnamefont
  {Heuer}},\ }\bibfield  {title} {\enquote {\bibinfo {title} {Theoretical
  description of ion conduction in disordered systems: From linear to nonlinear
  response},}\ }\href@noop {} {\bibfield  {journal} {\bibinfo  {journal}
  {Zeitschrift f{\"u}r Physikalische Chemie}\ }\textbf {\bibinfo {volume}
  {224}},\ \bibinfo {pages} {1855--1889} (\bibinfo {year} {2010})}\BibitemShut
  {NoStop}%
\bibitem [{\citenamefont {{Van Der Spoel}}\ \emph {et~al.}(2005)\citenamefont
  {{Van Der Spoel}}, \citenamefont {Lindahl}, \citenamefont {Hess},
  \citenamefont {Groenhof}, \citenamefont {Mark},\ and\ \citenamefont
  {Berendsen}}]{VanDerSpoel2005}%
  \BibitemOpen
  \bibfield  {author} {\bibinfo {author} {\bibfnamefont {D.}~\bibnamefont {{Van
  Der Spoel}}}, \bibinfo {author} {\bibfnamefont {E.}~\bibnamefont {Lindahl}},
  \bibinfo {author} {\bibfnamefont {B.}~\bibnamefont {Hess}}, \bibinfo {author}
  {\bibfnamefont {G.}~\bibnamefont {Groenhof}}, \bibinfo {author}
  {\bibfnamefont {A.~E.}\ \bibnamefont {Mark}}, \ and\ \bibinfo {author}
  {\bibfnamefont {H.~J.}\ \bibnamefont {Berendsen}},\ }\bibfield  {title}
  {\enquote {\bibinfo {title} {{GROMACS: Fast, flexible, and free}},}\ }\href
  {\doibase 10.1002/jcc.20291} {\bibfield  {journal} {\bibinfo  {journal}
  {Journal of Computational Chemistry}\ }\textbf {\bibinfo {volume} {26}},\
  \bibinfo {pages} {1701--1718} (\bibinfo {year} {2005})}\BibitemShut {NoStop}%
\bibitem [{\citenamefont {P{\'a}ll}\ \emph {et~al.}(2015)\citenamefont
  {P{\'a}ll}, \citenamefont {Abraham}, \citenamefont {Kutzner}, \citenamefont
  {Hess},\ and\ \citenamefont {Lindahl}}]{Pall2015}%
  \BibitemOpen
  \bibfield  {author} {\bibinfo {author} {\bibfnamefont {S.}~\bibnamefont
  {P{\'a}ll}}, \bibinfo {author} {\bibfnamefont {M.~J.}\ \bibnamefont
  {Abraham}}, \bibinfo {author} {\bibfnamefont {C.}~\bibnamefont {Kutzner}},
  \bibinfo {author} {\bibfnamefont {B.}~\bibnamefont {Hess}}, \ and\ \bibinfo
  {author} {\bibfnamefont {E.}~\bibnamefont {Lindahl}},\ }\bibfield  {title}
  {\enquote {\bibinfo {title} {Tackling exascale software challenges in
  molecular dynamics simulations with gromacs},}\ }in\ \href@noop {} {\emph
  {\bibinfo {booktitle} {Solving Software Challenges for Exascale}}},\ \bibinfo
  {editor} {edited by\ \bibinfo {editor} {\bibfnamefont {S.}~\bibnamefont
  {Markidis}}\ and\ \bibinfo {editor} {\bibfnamefont {E.}~\bibnamefont
  {Laure}}}\ (\bibinfo  {publisher} {Springer International Publishing},\
  \bibinfo {address} {Cham},\ \bibinfo {year} {2015})\ pp.\ \bibinfo {pages}
  {3--27}\BibitemShut {NoStop}%
\bibitem [{\citenamefont {Abraham}\ \emph {et~al.}(2015)\citenamefont
  {Abraham}, \citenamefont {Murtola}, \citenamefont {Schulz}, \citenamefont
  {P{\'{a}}ll}, \citenamefont {Smith}, \citenamefont {Hess},\ and\
  \citenamefont {Lindah}}]{Abraham2015}%
  \BibitemOpen
  \bibfield  {author} {\bibinfo {author} {\bibfnamefont {M.~J.}\ \bibnamefont
  {Abraham}}, \bibinfo {author} {\bibfnamefont {T.}~\bibnamefont {Murtola}},
  \bibinfo {author} {\bibfnamefont {R.}~\bibnamefont {Schulz}}, \bibinfo
  {author} {\bibfnamefont {S.}~\bibnamefont {P{\'{a}}ll}}, \bibinfo {author}
  {\bibfnamefont {J.~C.}\ \bibnamefont {Smith}}, \bibinfo {author}
  {\bibfnamefont {B.}~\bibnamefont {Hess}}, \ and\ \bibinfo {author}
  {\bibfnamefont {E.}~\bibnamefont {Lindah}},\ }\bibfield  {title} {\enquote
  {\bibinfo {title} {{Gromacs: High performance molecular simulations through
  multi-level parallelism from laptops to supercomputers}},}\ }\href {\doibase
  10.1016/j.softx.2015.06.001} {\bibfield  {journal} {\bibinfo  {journal}
  {SoftwareX}\ }\textbf {\bibinfo {volume} {1-2}},\ \bibinfo {pages} {19--25}
  (\bibinfo {year} {2015})}\BibitemShut {NoStop}%
\bibitem [{\citenamefont {Berendsen}, \citenamefont {van~der Spoel},\ and\
  \citenamefont {van Drunen}(1995)}]{Berendsen1995}%
  \BibitemOpen
  \bibfield  {author} {\bibinfo {author} {\bibfnamefont {H.~J.}\ \bibnamefont
  {Berendsen}}, \bibinfo {author} {\bibfnamefont {D.}~\bibnamefont {van~der
  Spoel}}, \ and\ \bibinfo {author} {\bibfnamefont {R.}~\bibnamefont {van
  Drunen}},\ }\bibfield  {title} {\enquote {\bibinfo {title} {{GROMACS: A
  message-passing parallel molecular dynamics implementation}},}\ }\href
  {\doibase 10.1016/0010-4655(95)00042-E} {\bibfield  {journal} {\bibinfo
  {journal} {Computer Physics Communications}\ }\textbf {\bibinfo {volume}
  {91}},\ \bibinfo {pages} {43--56} (\bibinfo {year} {1995})}\BibitemShut
  {NoStop}%
\bibitem [{\citenamefont {Gouveia}\ \emph {et~al.}(2017)\citenamefont
  {Gouveia}, \citenamefont {Bernardes}, \citenamefont {Tom{\'e}}, \citenamefont
  {Lozinskaya}, \citenamefont {Vygodskii}, \citenamefont {Shaplov},
  \citenamefont {Lopes},\ and\ \citenamefont {Marrucho}}]{gouveia2017ionic}%
  \BibitemOpen
  \bibfield  {author} {\bibinfo {author} {\bibfnamefont {A.~S.}\ \bibnamefont
  {Gouveia}}, \bibinfo {author} {\bibfnamefont {C.~E.}\ \bibnamefont
  {Bernardes}}, \bibinfo {author} {\bibfnamefont {L.~C.}\ \bibnamefont
  {Tom{\'e}}}, \bibinfo {author} {\bibfnamefont {E.~I.}\ \bibnamefont
  {Lozinskaya}}, \bibinfo {author} {\bibfnamefont {Y.~S.}\ \bibnamefont
  {Vygodskii}}, \bibinfo {author} {\bibfnamefont {A.~S.}\ \bibnamefont
  {Shaplov}}, \bibinfo {author} {\bibfnamefont {J.~N.~C.}\ \bibnamefont
  {Lopes}}, \ and\ \bibinfo {author} {\bibfnamefont {I.~M.}\ \bibnamefont
  {Marrucho}},\ }\bibfield  {title} {\enquote {\bibinfo {title} {Ionic liquids
  with anions based on fluorosulfonyl derivatives: from asymmetrical
  substitutions to a consistent force field model},}\ }\href@noop {} {\bibfield
   {journal} {\bibinfo  {journal} {Physical Chemistry Chemical Physics}\
  }\textbf {\bibinfo {volume} {19}},\ \bibinfo {pages} {29617--29624} (\bibinfo
  {year} {2017})}\BibitemShut {NoStop}%
\bibitem [{\citenamefont {{Canongia Lopes}}\ and\ \citenamefont
  {P\'adua}(2012)}]{CanongiaLopes2012}%
  \BibitemOpen
  \bibfield  {author} {\bibinfo {author} {\bibfnamefont {J.~N.}\ \bibnamefont
  {{Canongia Lopes}}}\ and\ \bibinfo {author} {\bibfnamefont {A.~A.}\
  \bibnamefont {P\'adua}},\ }\bibfield  {title} {\enquote {\bibinfo {title}
  {{CL\&P: A generic and systematic force field for ionic liquids modeling}},}\
  }\href {\doibase 10.1007/s00214-012-1129-7} {\bibfield  {journal} {\bibinfo
  {journal} {Theoretical Chemistry Accounts}\ }\textbf {\bibinfo {volume}
  {131}},\ \bibinfo {pages} {1--11} (\bibinfo {year} {2012})}\BibitemShut
  {NoStop}%
\bibitem [{\citenamefont {{Canongia Lopes}}, \citenamefont {Deschamps},\ and\
  \citenamefont {P\'adua}(2004)}]{JoseN.CanongiaLopes2004}%
  \BibitemOpen
  \bibfield  {author} {\bibinfo {author} {\bibfnamefont {J.~N.}\ \bibnamefont
  {{Canongia Lopes}}}, \bibinfo {author} {\bibfnamefont {J.}~\bibnamefont
  {Deschamps}}, \ and\ \bibinfo {author} {\bibfnamefont {A.~A.~H.}\
  \bibnamefont {P\'adua}},\ }\bibfield  {title} {\enquote {\bibinfo {title}
  {{Modeling Ionic Liquids Using a Systematic All-Atom Force Field}},}\ }\href
  {\doibase 10.1021/jp0362133} {\bibfield  {journal} {\bibinfo  {journal} {The
  Journal of Physical Chemistry B}\ }\textbf {\bibinfo {volume} {108}},\
  \bibinfo {pages} {2038--2047} (\bibinfo {year} {2004})}\BibitemShut {NoStop}%
\bibitem [{\citenamefont {Lopes}\ and\ \citenamefont
  {P\'adua}(2004)}]{Lopes2004}%
  \BibitemOpen
  \bibfield  {author} {\bibinfo {author} {\bibfnamefont {J.~N.}\ \bibnamefont
  {Lopes}}\ and\ \bibinfo {author} {\bibfnamefont {A.~A.}\ \bibnamefont
  {P\'adua}},\ }\bibfield  {title} {\enquote {\bibinfo {title} {{Molecular
  force field for ionic liquids composed of triflate or bistriflylimide
  anions}},}\ }\href {\doibase 10.1021/jp0476545} {\bibfield  {journal}
  {\bibinfo  {journal} {Journal of Physical Chemistry B}\ }\textbf {\bibinfo
  {volume} {108}},\ \bibinfo {pages} {16893--16898} (\bibinfo {year}
  {2004})}\BibitemShut {NoStop}%
\bibitem [{\citenamefont {Shimizu}\ \emph {et~al.}(2010)\citenamefont
  {Shimizu}, \citenamefont {Almantariotis}, \citenamefont {{Costa Gomes}},
  \citenamefont {P{\'{a}}dua},\ and\ \citenamefont {{Canongia
  Lopes}}}]{Shimizu2010}%
  \BibitemOpen
  \bibfield  {author} {\bibinfo {author} {\bibfnamefont {K.}~\bibnamefont
  {Shimizu}}, \bibinfo {author} {\bibfnamefont {D.}~\bibnamefont
  {Almantariotis}}, \bibinfo {author} {\bibfnamefont {M.~F.}\ \bibnamefont
  {{Costa Gomes}}}, \bibinfo {author} {\bibfnamefont {A.~A.}\ \bibnamefont
  {P{\'{a}}dua}}, \ and\ \bibinfo {author} {\bibfnamefont {J.~N.}\ \bibnamefont
  {{Canongia Lopes}}},\ }\bibfield  {title} {\enquote {\bibinfo {title}
  {{Molecular force field for ionic liquids V: Hydroxyethylimidazolium,
  dimethoxy-2methylimidazolium, and fluoroalkylimidazolium cations and
  Bis(fluorosulfonyl)amide, perfluoroalkanesulfonylamide, and
  fluoroalkylfluorophosphate anions}},}\ }\href {\doibase 10.1021/jp9120468}
  {\bibfield  {journal} {\bibinfo  {journal} {Journal of Physical Chemistry B}\
  }\textbf {\bibinfo {volume} {114}},\ \bibinfo {pages} {3592--3600} (\bibinfo
  {year} {2010})}\BibitemShut {NoStop}%
\bibitem [{\citenamefont {Parrinello}\ and\ \citenamefont
  {Rahman}(1981)}]{Parrinello1981}%
  \BibitemOpen
  \bibfield  {author} {\bibinfo {author} {\bibfnamefont {M.}~\bibnamefont
  {Parrinello}}\ and\ \bibinfo {author} {\bibfnamefont {A.}~\bibnamefont
  {Rahman}},\ }\bibfield  {title} {\enquote {\bibinfo {title} {{Polymorphic
  transitions in single crystals: A new molecular dynamics method}},}\ }\href
  {\doibase 10.1063/1.328693} {\bibfield  {journal} {\bibinfo  {journal}
  {Journal of Applied Physics}\ }\textbf {\bibinfo {volume} {52}},\ \bibinfo
  {pages} {7182--7190} (\bibinfo {year} {1981})}\BibitemShut {NoStop}%
\bibitem [{\citenamefont {Nos{\'{e}}}\ and\ \citenamefont
  {Klein}(1983)}]{Nose1983}%
  \BibitemOpen
  \bibfield  {author} {\bibinfo {author} {\bibfnamefont {S.}~\bibnamefont
  {Nos{\'{e}}}}\ and\ \bibinfo {author} {\bibfnamefont {M.}~\bibnamefont
  {Klein}},\ }\bibfield  {title} {\enquote {\bibinfo {title} {{Constant
  pressure molecular dynamics for molecular systems}},}\ }\href {\doibase
  10.1080/00268978300102851} {\bibfield  {journal} {\bibinfo  {journal}
  {Molecular Physics}\ }\textbf {\bibinfo {volume} {50}},\ \bibinfo {pages}
  {1055--1076} (\bibinfo {year} {1983})}\BibitemShut {NoStop}%
\bibitem [{\citenamefont {Nos{\'{e}}}(1984)}]{Nose1984}%
  \BibitemOpen
  \bibfield  {author} {\bibinfo {author} {\bibfnamefont {S.}~\bibnamefont
  {Nos{\'{e}}}},\ }\bibfield  {title} {\enquote {\bibinfo {title} {{A molecular
  dynamics method for simulations in the canonical ensemble}},}\ }\href
  {\doibase 10.1080/00268978400101201} {\bibfield  {journal} {\bibinfo
  {journal} {Molecular Physics}\ }\textbf {\bibinfo {volume} {52}},\ \bibinfo
  {pages} {255--268} (\bibinfo {year} {1984})}\BibitemShut {NoStop}%
\bibitem [{\citenamefont {Hoover}(1985)}]{Hoover1985}%
  \BibitemOpen
  \bibfield  {author} {\bibinfo {author} {\bibfnamefont {W.~G.}\ \bibnamefont
  {Hoover}},\ }\bibfield  {title} {\enquote {\bibinfo {title} {{Canonical
  dynamics: Equilibrium phase-space distributions}},}\ }\href {\doibase
  10.1103/PhysRevA.31.1695} {\bibfield  {journal} {\bibinfo  {journal}
  {Physical Review A}\ }\textbf {\bibinfo {volume} {31}},\ \bibinfo {pages}
  {1695--1697} (\bibinfo {year} {1985})}\BibitemShut {NoStop}%
\bibitem [{\citenamefont {Gorecki}\ \emph {et~al.}(1995)\citenamefont
  {Gorecki}, \citenamefont {Jeannin}, \citenamefont {Belorizky}, \citenamefont
  {Roux},\ and\ \citenamefont {Armand}}]{gorecki1995physical}%
  \BibitemOpen
  \bibfield  {author} {\bibinfo {author} {\bibfnamefont {W.}~\bibnamefont
  {Gorecki}}, \bibinfo {author} {\bibfnamefont {M.}~\bibnamefont {Jeannin}},
  \bibinfo {author} {\bibfnamefont {E.}~\bibnamefont {Belorizky}}, \bibinfo
  {author} {\bibfnamefont {C.}~\bibnamefont {Roux}}, \ and\ \bibinfo {author}
  {\bibfnamefont {M.}~\bibnamefont {Armand}},\ }\bibfield  {title} {\enquote
  {\bibinfo {title} {Physical properties of solid polymer electrolyte peo
  (litfsi) complexes},}\ }\href@noop {} {\bibfield  {journal} {\bibinfo
  {journal} {Journal of Physics: Condensed Matter}\ }\textbf {\bibinfo {volume}
  {7}},\ \bibinfo {pages} {6823} (\bibinfo {year} {1995})}\BibitemShut
  {NoStop}%
\bibitem [{\citenamefont {Shell}, \citenamefont {Debenedetti},\ and\
  \citenamefont {Stillinger}(2005)}]{Shell_2005}%
  \BibitemOpen
  \bibfield  {author} {\bibinfo {author} {\bibfnamefont {M.~S.}\ \bibnamefont
  {Shell}}, \bibinfo {author} {\bibfnamefont {P.~G.}\ \bibnamefont
  {Debenedetti}}, \ and\ \bibinfo {author} {\bibfnamefont {F.~H.}\ \bibnamefont
  {Stillinger}},\ }\bibfield  {title} {\enquote {\bibinfo {title} {Dynamic
  heterogeneity and non-gaussian behaviour in a model supercooled liquid},}\
  }\href {\doibase 10.1088/0953-8984/17/49/002} {\bibfield  {journal} {\bibinfo
   {journal} {Journal of Physics: Condensed Matter}\ }\textbf {\bibinfo
  {volume} {17}},\ \bibinfo {pages} {S4035} (\bibinfo {year}
  {2005})}\BibitemShut {NoStop}%
\bibitem [{\citenamefont {Balbuena}, \citenamefont {Frechero},\ and\
  \citenamefont {Montani}(2014)}]{BALBUENA2014124}%
  \BibitemOpen
  \bibfield  {author} {\bibinfo {author} {\bibfnamefont {C.}~\bibnamefont
  {Balbuena}}, \bibinfo {author} {\bibfnamefont {M.}~\bibnamefont {Frechero}},
  \ and\ \bibinfo {author} {\bibfnamefont {R.}~\bibnamefont {Montani}},\
  }\bibfield  {title} {\enquote {\bibinfo {title} {Channel diffusion in a
  lithium–potassium metasilicate glass using the isoconfigurational ensemble:
  Towards a scenario for the mixed alkali effect},}\ }\href {\doibase
  https://doi.org/10.1016/j.jnoncrysol.2014.09.001} {\bibfield  {journal}
  {\bibinfo  {journal} {Journal of Non-Crystalline Solids}\ }\textbf {\bibinfo
  {volume} {405}},\ \bibinfo {pages} {124--128} (\bibinfo {year}
  {2014})}\BibitemShut {NoStop}%
\end{thebibliography}%

\end{document}